\documentclass[12pt]{article}
\usepackage[english]{babel}
\usepackage{amsmath}
\usepackage{cite}
\usepackage[dvips]{graphicx}
\usepackage{amsfonts}
\usepackage{amssymb}
\begin{document}

\title{{\bf{}Programming Realization of Symbolic Computations for
Non-linear Commutator Superalgebras over the Heisenberg--Weyl
Superalgebra: Data Structures and Processing Methods}}

\author{\sc A. Kuleshov${}^{a}$\thanks{ksv1986@sibmail.com
\hspace{1.5cm} ${}^{\dagger}$ reshet@ispms.tsc.ru}, A.A.
Reshetnyak$^{b\dagger}$ \\[0.5cm]
\it ${}^a$Department on Informatics,\\
\it Tomsk State  University,\\
\it Tomsk 634050, Russia\\[0.3cm]
\it ${}^{b}$Laboratory of Non-Linear Media Physics,\\
\it  Institute of  Strength Physics
and Materials Science,\\
\it Tomsk 634021, Russia}
\date{}

\maketitle
\thispagestyle{empty}

\begin{abstract}

We suggest a programming realization of an algorithm for a
verification of a given set of algebraic relations in the form of
a supercommutator multiplication table for the Verma module, which
is constructed according to a generalized Cartan procedure for a
quadratic superalgebra and whose elements are realized as a formal
power series with respect to non-commuting elements. To this end,
we propose an algebraic procedure of Verma module construction and
its realization in terms of non-commuting creation and
annihilation operators of a given Heisenberg--Weyl superalgebra.
In doing so, we set up a problem which naturally arises within a
Lagrangian description of higher-spin fields in anti-de-Sitter
(AdS) spaces: to verify the fact that the resulting Verma module
elements obey the given commutator multiplication for the original
non-linear superalgebra. The problem setting is based on a
restricted principle of mathematical induction, in powers of
inverse squared radius of the AdS-space. For a construction of an
algorithm resolving this problem, we use a two-level data model
within the object-oriented approach, which is realized on a basis
of the programming language C\#. The first level, the so-called
\emph{basic model of superalgebra}, describes a set of operations
to be realized as symbolic computations for arbitrary
finite-dimensional associative superalgebras. The second level
serves to realize a specific representation of non-linear
commutator superalgebra elements, and specifies the peculiarities
of commutation operations for the elements of a specific
superalgebra $\mathcal{A}$, as well as the ordering of creation
$f^+, b_i^+$ and annihilation $f, b_i$, $i=1,2$, operators in
products which determine supercommutators $[a,b\} $, $a,b \in
\mathcal{A}$, to be verified. The program allows one to consider
objects (of a less general nature than non-linear commutator
superalgebras) that fall under the class of so-called
$GR$-algebras, for whose treatment one widely uses the module
\emph{Plural} of the system \emph{Singular} of symbolic
computations for polynomials.
\end{abstract}



\section{Introduction}

The problem of treatment of algebraic structures more general than
Lie algebras \cite{BarutRonchka} and superalgebras
\cite{LeitesSUSY}, equivalent, in fact, to matrix algebras and
superalgebras, is a relatively recent issue in the area of
Theoretical Physics and Pure and Applied Mathematics; for a review
of notions on non-linear algebras, see the textbook
\cite{Morozov}. Mathematically, this trend gains its motivation
from the study of nonlinear algebras and superalgebras, such as
$W$-algebras \cite{Walgebras}, whereas from the physical viewpoint
it is due to an intensive application of nonlinear algebraic
structures in High Energy Physics, in particular, within the
theory of strings and superstrings \cite{strings} and the related
Higher Spin Field Theory; for a review see \cite{reviews}.
Field-theoretical models of higher-spin (HS) fields in
constant-curvature spaces (Minkowski, de Sitter, anti-de-Sitter)
related to the hope of detection (perhaps in view of the expected
launch of LHC), at a level of energy higher than the level
presently accessible to physical laboratories, of new kinds of
interactions and particles which must be part of superstring
spectrum. It should be noted that the choice of the anti-de-Sitter
(AdS) space presents, first of all, the simplest non-trivial
background providing a consistent propagation of free \cite{FF}
and interacting HS fields, since the radius of the AdS space
ensures the presence of a natural dimensional parameter for an
accommodation of compatible self-interactions \cite{Fradkin,
Sezgin}. Second, the (A)dS space is the most adequate model for a
description of space-time corresponding to the Universe, in view
of the modern data \cite{accelerated} on its accelerated
expansion. Third, HS fields in the AdS space are closely related
to the tensionless limit of superstring theory on the $AdS_5
\times S_5$ Ramond--Ramond background \cite{Heslop,Bonelli1} and
the conformal $\mathcal{N} = 4$ SYM theory in the context of the
AdS/CFT correspondence \cite{Maldacena}.

For a quantum description of an HS field in the AdS$_d$-spaces
within conventional Quantum Field Theory, it is necessary to
construct its gauge-invariant Lagrangian description, which
includes a determination of the action functional and of the set
of reducible gauge symmetries \cite{Weinberg,GT}; for the
pioneering works on this problem, see for instance \cite{FSingh}.
Among different methods\footnote{The light-cone formalism
\cite{Metsaev2}, Vasiliev's frame-like formalism \cite{Vasiliev,
Alkalaev,Grigoriev} using the unfolded approach \cite{unfolded},
Fronsdal's formalism \cite{Fronsdal}, the constrained
\cite{ZinovievbosAdS0211233} and unconstrained \cite{Sagnotti},
metric-like formalism.} which allow one to solve this problem, an
especially outstanding one is the \emph{BFV--BRST approach},
inspired by Witten's String Field Theory \cite{strings}, and based
on a special global BRST symmetry \cite{BRST} and on the BFV
method \cite{BFV}, realizing this symmetry within the Hamiltonian
description of dynamical systems with constraints.

For the purpose of this work, it is appropriate to mention that
the central object of the BFV--BRST approach, the BFV--BRST
operator, is constructed, in the case of the AdS$_d$-space, with
respect to a non-linear (super)algebra
$\mathcal{A}_{c}(Y(1),AdS_d)$\footnote{Here, following to Ref.
\cite{Resh08122329} and in view of absence of classification and
generally-accepted terminology for nonlinear (super)algebras, we
suggest the notation  $\mathcal{A}(Y(k),AdS_d)$  for nonlinear
superalgebra of initial operators which correspond to half-integer
HS fields in AdS${}_d$ space subject to Young tableaux with $k$
rows, the same for nonlinear superalgebras of converted operators
$\mathcal{A}_{c}(Y(k),AdS_d)$  and of additional parts
$\mathcal{A}'(Y(k),AdS_d)$. The case of nonlinear algebras for
integer spin HS fields is labeled by means of subscript "b" in
$\mathcal{A}_{b}(Y(k),AdS_d)$, $\mathcal{A}_{b}'(Y(k),AdS_d)$,
$\mathcal{A}_{b{}c}(Y(k),AdS_d)$.}, $\mathcal{A}_{c}(Y(1),AdS_d)$
=$\mathcal{A}(Y(1),AdS_d)$ + $\mathcal{A}^{\prime}(Y(1),AdS_d)$,
for a (half-)integer spin ($s = n+\frac{1}{2}$, $n \in
\mathbb{N}_0$) $s, s\in \mathbb{N}_0$ subject to a Young tableaux
with one row. The operators $O_I$, $O_I = (o_I+o'_I)$, composing
this (super)algebra $O_I \in \mathcal{A}_{c}(Y(1),AdS_d)$ are
determined in a special Hilbert space $\mathcal{H}_c$,
$\mathcal{H}_c = \mathcal{H} \otimes \mathcal{H}'$, with respect
to differential algebraic relations which extract a field (tensor,
$s\in \mathbb{N}_0$, or spin-tensor, $s = n+\frac{1}{2}$) of given
mass $m$ and spin $s$ from the space of unitary irreducible
representation of the AdS group in the AdS$_d$ space.

Having restricted this paper by the case of a half-integer spin,
we note that the deduction of the superalgebra
$\mathcal{A}_{c}(Y(1),AdS_d)$ is based \cite{adsfermBKR} on the
construction of an auxiliary representation, called the Verma
module \cite{Dixmier}, for a quadratic superalgebra
$\mathcal{A}^{\prime}(Y(1),AdS_d)$, coinciding with the
superalgebra $\mathcal{A}(Y(1),AdS_d)$ in a flat space, $r=0$,
i.e., for the Lie superalgebra
$\mathcal{A}(Y(1),\mathbb{R}^{d-1,1})$. The Verma module provides
a correct number of physical degrees of freedom in a non-Abelian
conversion method \cite{conversion} and therefore ensures an
application of the BFV--BRST approach. While the problem of Verma
module construction is solved for Lie algebras \cite{Liealgebra}
and superalgebras \cite{Liesusy, mixfermi}, for the quadratic
\cite{0206027} operator algebra $\mathcal{A}_{b}(Y(1),AdS_d)$ used
in Ref.~\cite{bkradsbose} to construct a Lagrangian formulation
for bosonic HS fields in the AdS$_d$-space subject to $Y(1)$, the
corresponding problem for the non-linear superalgebra
$\mathcal{A}^{\prime}(Y(1),AdS_d)$ is yet unsolved, leaving the
correctness of the Lagrangian formulation of Ref.
\cite{adsfermBKR} questionable.

The principal goals of this paper are as follows:
\begin{enumerate}
    \item development of a method of constructing
    the Verma module $V_{\mathcal{A}'}$ for a non-linear superalgebra
    $\mathcal{A}^{\prime}(Y(1),AdS_d)$ on the basis of a generalized
    Cartan procedure;
    \item realization of the Verma
    module $V_{\mathcal{A}'}$ in terms of a formal power series in
    the degrees of non-supercommuting generating elements $b_i,b_i^+,f,f^+, i=1,2$
    of the Heisenberg--Weyl superalgebra, whose number coincides with those
    of negative $\{E^{-A}\}$ and positive $\{E^A\}$
    root vectors in a Cartan-like decomposition
    $\mathcal{A}^{\prime}(Y(1),AdS_d)$ = $\mathcal{E}^{-A} \oplus
    \mathcal{H}^{\hat{i}} \oplus \mathcal{E}^{A}$ (with Cartan subsuperalgebra $\mathcal{H}^{\hat{i}}$).
\end{enumerate}
In connection with a solution of these problems, there arises a
number of peculiarities, stipulated by the fact that the elements
of the Verma module $V_{\mathcal{A}'}$ are constructed with respect to
a given multiplication for the superalgebra
$\mathcal{A}^{\prime}(Y(1),AdS_d)$ in an indirect way:
\begin{itemize}
    \item
first, the Verma module $V_{\mathcal{A}'}$ is derived by means of
the Cartan procedure, and then it is realized as a formal power series
$o'_I(b_i,b_i^+,f,f^+)$; as a consequence, one needs a formal proof
of the fact that these operators $o'_I(b_i,b_i^+,f,f^+)$ actually
satisfy the table of supercommutator multiplication for
$\mathcal{A}^{\prime}(Y(1),AdS_d)$;
    \item in view of a sufficiently large number of basis elements, $l=9$, for
$\mathcal{A}^{\prime}$, which grows with the increasing
of the rows of the Young tableaux, so that for
$\mathcal{A}^{\prime}(Y(k),AdS_d)$ the number of basis elements is
equal to $2(1+k^2+\frac{5}{2}k)$, from the technical viewpoint the
problem of verifying the given algebraic relations is quite time-consuming.
\end{itemize}

Therefore, the next group of problems to be solved in this paper
is the following:
\begin{enumerate}
    \item[3.] finding a {\it formalized setting} of the problem of verifying
    the fact that the operators $o'_I(b_i,b_i^+$, $f,f^+)$ obey the given
    algebraic relations $\mathcal{A}^{\prime}(Y(1),AdS_d)$
    with the help of a restricted induction principle, in powers of
     the inverse squared radius $r$ of the AdS$_d$-space;
    \item[4.] realization, in a high-level programming language ($C\#$),
    of an algorithm of solving the above problem by using the techniques
    of symbolic calculations.
\end{enumerate}

From the viewpoint of mathematics and programming, the problem of
symbolic computations (the symbols here are the elements of the
Heisenberg--Weyl superalgebra $b_i,b_i^+,f,f^+$ and polynomials
constructed from them) with respect to non-linear superalgebras
has not been considered, and, to our knowledge (see, for instance,
\cite{refer1} and references therein), has not been realized as a
computer program. The particular case of the flat space
$\mathbb{R}^{1,d-1}$ for $r=0$ is an exception where such
well-known application packages as Maple, MathLab, MathCad,
Mathematica, etc., permit one to operate with the Lie superalgebra
$\mathcal{A}^{\prime}(Y(1), \mathbb{R}^{1,d-1})$, equivalent to
supermatrix algebras. In addition, formal power series
$o'_I(b_i,b_i^+,f,f^+)$ pass in this case to finite-order
polynomials of at most third degree with respect to
$b_i,b_i^+,f,f^+$, so that the solution of problem $3$ appears
quite trivial for a calculator. It should be noted that among the
programs being the most capable to work with symbolic
calculations one widely uses the module \emph{Plural}
\cite{Plural} of the system \emph{Singular} for symbolic
calculations of polynomials, which is intended for computations in
a class of non-commuting polynomial algebras. Left ideals and
modules over a given non-commutative $G$\emph{-algebra}
\cite{Galgebra}, so-called, $GR$-\emph{algebras}, are the basic
objects of calculations using \emph{Plural}. At the same time, the
case of the nonlinear superalgebra
$\mathcal{A}^{\prime}(Y(1),AdS_d)$ under consideration has a
number of supercommutator relations that cannot be realized within
the class of $G$-algebras and therefore in \emph{Plural} as well.

The paper is organized as follows. In Section~\ref{NLnSualg}, we
introduce necessary algebraic definitions, examine a special
nonlinear operator superalgebra
$\mathcal{A}^{\prime}(Y(1),AdS_d)$, whose algebraic relations were
obtained in Ref.~\cite{adsfermBKR}, explicitly construct the Verma
module $V_{\mathcal{A}'}$, find a realization of
$V_{\mathcal{A}'}$ in terms of a formal power series in
noncommuting elements (symbols) of the Heisenberg--Weyl
superalgebra $A_{1,2}$, and set up a formalized representation for
$\mathcal{A}^{\prime}(Y(1),AdS_d)$. In Section~\ref{Programmreal},
we consider in detail the elements of a programming realization
using $C\#$ to solve the formalized setting of the problem on the
basis of a two-level model for a representation of the Verma
module for a nonlinear superalgebra, which includes a so-called
\emph{basic model of superalgebra} and \emph{model of polynomial
superalgebra}. We apply the developed program to a verification of
the required algebraic relations for the superalgebra
$\mathcal{A}^{\prime}(Y(1),AdS_d)$ in Section~\ref{Programm}. In
Section~\ref{ConclusionP}, we summarize the results of the paper
and discuss the perspectives of applying the program.

\section{Non-Linear Superalgebras }\label{NLnSualg}

In this section, we introduce the definitions of a non-linear
superalgebra with respect to commutator multiplication and study a
number of its algebraic properties for a solution of the basic
algebraic problems for a special operator superalgebra. We then
use our construction to develop a problem setting in order to
fulfill a program verification of the fact that a given oscillator
realization of the above superalgebra actually satisfies
a given multiplication table.

\subsection{Basic definitions and algebraic constructions}\label{Basicdef}

Let $\mathbb{K}$ be a field and  $\mathcal{A} = \{e,o_I\}$, $I \in
\Delta$ be an associative $\mathbb{K}$-superalgebra with unity $e$
and a basis $\{e, o_I\}$, being a two-side module over a Grassmann
algebra $\Lambda = \{\alpha_k\}$, $k \in X$, where $\Delta$ and
$X$ are independent  finite or infinite sets of indices.

\textbf{Definition 1.} \emph{Associative $\mathbb{K}$-superalgebra
$\mathcal{A}$ over $\Lambda$ is called a \textbf{non-linear
Lie-type superalgebra}\footnote{Of course, a non-linear Lie-type
superalgebra $\mathcal{A}$ corresponding to the case of an
associative superalgebra $\mathcal{A}$ does not contain the
unity.} if there exists a two-place operation $[\ ,\ \}$
satisfying the following conditions for $I,J,K \in \Delta$, $k,l
\in X$:}
     \begin{eqnarray}
        \mathbf{- Non-linearity} && [o_I, o_J\}  =
        F^{K}_{IJ}(o)o_{K},\
        F^{K}_{IJ} = f^{(1)K}_{IJ} + \sum_{n=2}^\infty
        f^{(n)K_1...K_{n-1}K}_{IJ}\prod_{i=1}^{n-1} o_{K_i};
        \label{nly} \\
        \mathbf{- Antisymmetry\ }  &&  [o_I, o_J\} = - (-1)^{\varepsilon_I\varepsilon_J}
        [o_J, o_I\}; \label{asm}\\
        \mathbf{- \Lambda-bilinearity\ }  &&  [\alpha_ko_I + \alpha_lo_J, o_K\}=
        \alpha_k[o_I,o_K\}
        + \alpha_l[o_J, o_K\} , \nonumber\\
   \phantom{\mathbf{- \Lambda-bilinearity}} &&
    [o_I, \alpha_ko_J + \alpha_lo_K\} =
(-1)^{\varepsilon_I\varepsilon_k}\alpha_k[o_I, o_J\} +
(-1)^{\varepsilon_I\varepsilon_l}\alpha_l[o_I, o_K\}; \label{bily}
\\
         \mathbf{- Leibnitz\ rule\ \ \ } && [o_I, o_J  o_K\} = [o_I, o_J\} o_K + (-1)^{
  \varepsilon_I\varepsilon_J}o_J [o_I, o_K\}    \label{LeiR}\\
         \mathbf{- Jacobi\ identity} && (-1)^{\varepsilon_I\varepsilon_K}
         [o_I, [o_J, o_K\}\} + cycl.perm. (I, J, K) = 0, \label{Jid}
    \end{eqnarray}
\emph{where we suppose summation with respect to repeated indices
$K_1,...,K_{n-1},K$ in (\ref{nly}); the quantities $f^{(1)K}_{IJ},
f^{(n)K_1...K_{n-1}K}_{IJ} \in \Lambda$ obey the antisymmetry
properties }\begin{eqnarray} \bigl(F^{K}_{IJ}(o), f^{(1)K}_{IJ},
f^{(n)K_1...K_{n-1}K}_{IJ}\bigr)  = -
(-1)^{\varepsilon_I\varepsilon_J}\bigl(F^{K}_{JI}(o),
f^{(1)K}_{JI}, f^{(n)K_1...K_{n-1}K}_{JI}\bigr),
\label{antproperty}
\end{eqnarray}
\emph{and $\varepsilon_I, \varepsilon_k$ are the Grassmann parities of
the elements $o_I$, $\alpha_k$, $(\varepsilon_I, \varepsilon_k)$
$\equiv$ $(\varepsilon(o_I), \varepsilon(\alpha_k))$, being equal
respectively to 0 and 1 for even and odd $o_I,\alpha_k$ with
respect to a given $\mathbb{Z}_2$-grading in $\mathcal{A}$}.

It should be noted, first, that the property (\ref{LeiR}) presents
the compatibility of the associative and Lie-type multiplications
in $\mathcal{A}$. Second, a typical example of a non-linear
Lie-type superalgebra is the classical analogue of
finite-dimensional non-linear superalgebras \cite{Lavrov1} where
the operation $[\ ,\ \}$ is the Poisson superbracket $\{\ ,\
\}$.

\textbf{Definition 2.} \emph{A non-linear Lie-type superalgebra
$\mathcal{A}$ is called a \textbf{non-linear commutator
superalgebra} if the two-place operation $[\ ,\ \}$ is realized as
a supercommutator for any $A, B \in \mathcal{A}$ with definite
Grassmann gradings $\varepsilon(A), \varepsilon(B)$}
\begin{equation}\label{supercomm}
 [A, B\} = AB - (-1)^{\varepsilon(A)\varepsilon(B)}BA.
\end{equation}

\textbf{Definition 3.} \emph{A non-linear commutator (Lie-type)
superalgebra $\mathcal{A}$ is called a \textbf{polynomial (Lie
type) superalgebra of order $n$}, $n \in \mathbb{N}$, if
decomposition (\ref{nly}) obeys the following condition:}
\begin{equation}\label{polynom}
f^{(n)KK_1...K_n}_{IJ} \neq 0,\ and\
f^{(k)KK_1...K_nK_{n+1}...K_k}_{IJ}=0,\ k>n, k\in \mathbb{N}.
\end{equation}

\noindent \textbf{Corollary 1.} Polynomial superalgebras of order
$1$, $2$ correspond to Lie superalgebras \cite{LeitesSUSY} and
quadratic superalgebras \cite{adsfermBKR}, such as superconformal
algebras, extending the case of quadratic algebras
\cite{quadratic, Walgebras}.

It should be noted that within the class of polynomial algebras
and superalgebras of definite order $k$ there exist
 superalgebras \cite{parafermion}, \cite{parafermion1} with so called
parasupersymmetry and
 superalgebras with only parabosonic elements \cite{paraboson0}, the ones with non-linear realization of the
 supersymmetry used
 in the framework of mechanics, in description of
 Aharonov-Bohm effect \cite{paraboson1}, \cite{paraboson2}.

It is interesting to observe the structure of the following
relations for a non-linear Lie type superalgebra, starting from
the resolution of the Jacobi identity (\ref{Jid}) for the elements
$\{o_I\}$. In doing so, we may follow two ways: first, a purely
\emph{algebraic approach}, and, second, a more general
\emph{gauge-inspired approach} \cite{BFV,
algnonlnalgebraHenneaux}. For instance, in the case of a
supercommutative quadratic Lie-type superalgebra (which can be
considered as a generalization of a so-called \emph{Poisson $L-T$
algebra} \cite{9312183Henneaux} to the case of a superalgebra, if
$[\ ,\ \}$ is a Poisson bracket realized in a corresponding phase
space), we may obtain two sets of relations which present a
solution of the Jacobi identity (\ref{Jid}):
\begin{eqnarray}
{}&& \Bigl\{(-1)^{\varepsilon_I\varepsilon_K}\Bigl[
   f^{(1)K_1}_{IJ}f^{(1)L_3}_{K_1K} +
   \bigl(f^{(1)K_1}_{IJ}f^{(2)L_2L_3}_{K_1K} + (-1)^{\varepsilon_K
   \varepsilon_{L_3}}
   f^{(2)K_1L_3}_{IJ}f^{(1)L_2}_{K_1K}  \nonumber \\
{}&& + (-1)^{\varepsilon_{L_2}(\varepsilon_{L_3}+\varepsilon_{K_1}
+ \varepsilon_K)}
   f^{(2)L_2K_1}_{IJ}f^{(1)L_3}_{K_1K}\bigr)o_{L_2}+ \bigl((-1)^{\varepsilon_K
   \varepsilon_{L_3}}f^{(2)K_1L_3}_{IJ}f^{(2)L_1L_2}_{K_1K}\nonumber \\
{}&&  +
   (-1)^{\varepsilon_{L_1}(\varepsilon_{L_2}+ \varepsilon_{L_3} +
   \varepsilon_{K_1}+\varepsilon_K)}f^{(2)L_1K_1}_{IJ}f^{(2)L_2L_3}_{K_1K}
   \bigr)o_{L_1}o_{L_2}\Bigr] + cycl.perm. (I, J,
   K)\Bigr\}o_{L_3}=0, \label{explJacob}
\end{eqnarray}
in the algebraic approach, with the use of the obvious symmetry
property for $f^{(2)K_1K_2}_{IJ}$, $f^{(2)K_1K_2}_{IJ} =
(-1)^{\varepsilon_{K_1}\varepsilon_{K_2}}f^{(2)K_2K_1}_{IJ}$
following from supercommutativity of $o_{K_1},
o_{K_2}$\footnote{i.e. from the property: $o_{K_1} o_{K_2} =
\frac{1}{2}\bigl(o_{K_1} o_{K_2}
+(-1)^{\varepsilon_{K_1}\varepsilon_{K_2}} o_{K_2} o_{K_1}\bigr)$.
For a similar form of Jacobi identities see the paper
\cite{Lavrov1}.},
\begin{eqnarray}\label{resalgJid1}
   &&
   (-1)^{\varepsilon_I\varepsilon_K}
   f^{(1)K_1}_{IJ}f^{(1)L_3}_{K_1K}+ cycl.perm. (I, J,
   K)= 0,   \\
   && (-1)^{\varepsilon_I\varepsilon_K}
   \bigl[f^{(1)K_1}_{IJ}f^{(2)L_2L_3}_{K_1K} + (-1)^{\varepsilon_K
   \varepsilon_{L_3}}
   f^{(2)K_1L_3}_{IJ}f^{(1)L_2}_{K_1K} \nonumber\\
   && +
   (-1)^{\varepsilon_{L_2}(\varepsilon_{L_3}+ \varepsilon_K)}
   f^{(2)K_1L_2}_{IJ}f^{(1)L_3}_{K_1K}\bigr] + cycl.perm.(I,J,K) =0,
\label{resalgJid2}
 \\
   && X^{L_3L_1L_2}_{IJK} =(-1)^{\varepsilon_I\varepsilon_K}\Bigl\{ \bigl[(-1)^{
   \varepsilon_{L_1}(\varepsilon_{L_2}+\varepsilon_K)}
   f^{(2)K_1L_1}_{IJ}
   f^{(2)L_2L_3}_{K_1K} + cycl.perm. (L_1, L_2,
   L_3)\bigr]\Bigr\}\nonumber\\
   &&\qquad  + cycl.perm. (I, J, K)=0
   \label{resalgJid3}
   .
\end{eqnarray}
Whereas in the gauge-inspired approach there exist third-order
structure functions $F_{IJK}^{L_1L_2}(o)$ which satisfy the
properties
\begin{eqnarray}
  F_{IJK}^{L_1L_2}(o) &=& F_{IJK}^{(0)L_1L_2}+F_{IJK}^{(1)L_1L_2; M}o_M,
  \label{3strfunc}\\
  F_{IJK}^{L_1L_2} &=& -(-1)^{\varepsilon_I\varepsilon_J}F_{JIK}^{
  L_1L_2} = -(-1)^{\varepsilon_J\varepsilon_K}F_{IKJ}^{
  L_1L_2} = -(-1)^{\varepsilon_{L_1}\varepsilon_{L_2}}
  F_{IJK}^{L_2L_1}, \label{symmpropertyF3}
\end{eqnarray}
such that the relations which totally resolve the Jacobi identity
contain not only the standard Lie equation for structure
constants $f^{(1)K_1}_{IJ}$ (\ref{resalgJid1}) but, with a restriction
for $f^{(2)K_1K_2}_{IJ}$ given by Eq. (\ref{resalgJid2}),
also new relations:
\begin{eqnarray}
   && Z^{L_3L_1L_2}_{IJK} =
\Bigl[(-1)^{\varepsilon_I\varepsilon_K}\Bigl\{
   (-1)^{
   \varepsilon_{L_3}\varepsilon_K}
   f^{(2)K_1L_3}_{IJ}
   f^{(2)L_1L_2}_{K_1K} + \frac{1}{2}\bigl[(-1)^{\varepsilon_{L_1}(\varepsilon_{L_2}+ \varepsilon_{L_3} +
   \varepsilon_{K_1}+\varepsilon_K)}f^{(2)L_1K_1}_{IJ}f^{(2)L_2L_3}_{K_1K}  \nonumber\\
   &&  \qquad+
   (L_1 \longleftrightarrow L_2)\bigr]
   \Bigr\}+ cycl.perm. (I, J,
   K)\Bigr] - F_{IJK}^{(1)L_2L_3; L_1}=0
   \label{resgauJid3}
   .
\end{eqnarray}
The generalized symmetry property of the terms to be quadratic
in $o_{L_2}o_{L_3}$ in (\ref{explJacob}) with respect to upper
indices $({L_1},{L_2})$ leads to the identical vanishing of the
quantities $F_{IJK}^{(0)L_2L_3}$ due to relations
(\ref{symmpropertyF3}) in the case of a supercommutative superalgebra,
whereas the terms being cubic with respect to $o_{L_1}o_{L_2}o_{L_3}$
in (\ref{Jid}) may be only generalized-symmetric with respect to
$o_{L_1}o_{L_2}$, so that the set of quantities
$F_{IJK}^{(1)L_2L_3; L_1}$ contains the terms generalized-symmetric
with respect to a permutation of $(L_1, L_2)$.

The vanishing of the terms being generalized-antisymmetric with
respect to permutations $(L_1,L_3), (L_2,L_3)$ in Eqs.
(\ref{resgauJid3}), which means the vanishing of the quantities
$F_{IJK}^{(1)L_2L_3; L_1}$ as well, reduces (\ref{resgauJid3}) to
the relation (\ref{resalgJid3}) of the algebraic approach. The
quantities $F_{IJK}^{(1)L_2L_3; L_1}$:
\begin{equation}\label{explformF3}
F_{IJK}^{L_2L_3}(o) = F_{IJK}^{(1)L_2L_3; L_1}o_{L_1}\texttt{ for
}F_{IJK}^{(0)L_2L_3} = 0,
\end{equation}
 are generally not arbitrary and their
form is controlled by higher structure relations; see
\cite{algnonlnalgebraHenneaux} for details.

In obtaining the Jacobi identities, we only use
properties (\ref{nly})--(\ref{LeiR}) and take into account that
Eqs. (\ref{explJacob}) by themselves are valid for an arbitrary
non-linear Lie-type superalgebra without the requirement of
a supercommutativity for the usual multiplication in $\mathcal{A}$.
Of course, in the latter case  relations (\ref{resalgJid2}),
(\ref{resalgJid3}) [and (\ref{resgauJid3})] are not valid since
they have been obtained, first, with help of symmetrization with
respect to the free indices $(L_1, L_2,
   L_3)$ [and $(L_1, L_2)$], which no longer takes place, second,
   due to $F_{IJK}^{(0)L_2L_3} \neq 0$, and, third, because the former relations
   (\ref{resalgJid2}), (\ref{resalgJid3}) have been obtained from a more restrictive
   requirement of the vanishing of all the coefficients in front of algebraically
   independent symmetric monomials $\{o_{L_1}$, $o_{L_1}o_{L_2}$, $o_{L_1}o_{L_2}
   o_{L_3}\}$ as in \cite{9312183Henneaux} for the non-linear algebras and as in \cite{Lavrov1} for non-linear superalgebras\footnote{A resolution
   of the Jacobi identities by the gauge-inspired approach does not
   require symmetrization over upper indices, and the corresponding form of
   the Jacobi identities for a non-linear commutator superalgebra
   can be found in Ref.~\cite{Resh08122329}, whereas a detailed study
   of the structure of supercommutative Lie-type and non-linear commutator
   superalgebras will be presented in Ref. \cite{nlnsuperRM}.}.

   As the additional note, we only mention that for the case of
   solutions of the Jacobi identities in the form given by the
   Eqs.(\ref{resalgJid1}), (\ref{resalgJid2}), (\ref{resgauJid3})
   with vanishing third-order structural coefficients $F_{IJK}^{(1)L_2L_3;
   L_1}$ (and absence of the fourth-order coefficients $F_{IJKL}^{L_3L_2L_1}(o)$) the structure of nilpotent BRST operator $Q$ for superalgebra in question corresponds
to the case of closed algebra as follows:
\begin{eqnarray}\label{BRSTcl}
    {Q}  = \mathcal{C}^I\bigl[{{o}}_I  + \textstyle\frac{1}{2}
    \mathcal{C}^{J}(f^{(1)P}_{JI}+
   f^{(2){KL}}_{JI}o_K)\mathcal{P}_{L}
    (-1)^{\varepsilon_{I}+\varepsilon_{P}}\bigr]
\end{eqnarray}
with conjugated ghost coordinates $\mathcal{C}^I$ and momenta
$\mathcal{P}_{I}$ of opposite Grassmann parities to ones of $o_I$.
As the result, the BRST operator (\ref{BRSTcl}) coincides with one
in \cite{Lavrov1}, but there are not additional quadratic
restrictions (given by Eqs.(50) in \cite{Lavrov1}) on non-linear
second-order coefficients $f^{(2){KL}}_{IJ}$ out of the
Eqs.(\ref{resgauJid3}). Indeed, the corresponding restrictions
[with except for cubic relations on $f^{(1){K}}_{IJ}$,
$f^{(2){KL}}_{IJ}$ interrelated with absence of
$F_{IJKL}^{L_3L_2L_1}(o)$] on $f^{(2){KL}}_{IJ}$ are naturally
encoded by the non generalized-symmetric parts of
Eqs.(\ref{resgauJid3}) with respect to pairs of indices
$(L_2,L_3)$ and $(L_1,L_3)$, which have the form of the Eqs.(50)
in \cite{Lavrov1}:
\begin{eqnarray}
   Y^{L_3L_1L_2}_{IJK} =  (-1)^{(\varepsilon_I+\varepsilon_{L_3})\varepsilon_K}
   f^{(2)K_1L_3}_{IJ}
   f^{(2)L_1L_2}_{K_1K} + cycl.perm. (I, J,
   K)=0
   \label{resgauJid4}
   .
\end{eqnarray}
As the consequence, the Jacobi identities (\ref{resgauJid3}) after
deduction of the relations (\ref{resgauJid4}) multiplied on
$\frac{1}{2}$ are reduced to ones obtained from algebraic approach
(\ref{resalgJid3}): $Z^{L_3L_1L_2}_{IJK} -
\frac{1}{2}Y^{L_3L_1L_2}_{IJK} = X^{L_3L_1L_2}_{s{}IJK}$.

\subsubsection{Verma module $V_{\mathcal{A}'}$ construction for superalgebra
$\mathcal{A}^{\prime}(Y(1),AdS_d)$}\label{VermaSUSY}

Let us remind that the non-linear commutator superalgebra
$\mathcal{A}^{\prime}(Y(1),AdS_d)$ is formed by the generating
elements $\{o'_I\}$, $I=\overline{1,9}$, which contain 3 odd
(fermionic) and 6 even (bosonic) quantities with respect to
the Grassmann parity $\varepsilon$,
 \begin{eqnarray}\label{numoperators}
(o'_1,o'_2, o'_3) = (t_0', t'_1, t_1^{\prime +}),\  (o'_4,o'_5,
..., o'_9) =(l_0', l'_1, l_1^{\prime +}, l'_2, l_2^{\prime +},
g_0');\ \varepsilon(o'_I)=\left\{\begin{array}{c}
                            1, I=1,2,3, \\
                             0, I=4,...,9\\
                          \end{array}\right.,
\end{eqnarray}
and whose commutator products (\ref{nly}) are defined by the
multiplication table~\ref{table'}, given for the first time
in Ref.~\cite{adsfermBKR},
\begin{table}
\begin{center}
\begin{tabular}{||c||r|r||r|r|r|r|r|r|r||}\hline\hline
     $[\; \downarrow, \rightarrow \}$    &$t'_0$&$t'_1$&
       $t^{\prime+}_1$&${l}'_0$&$l'_1$&$l^{\prime+}_1$&$l'_2$&
       $l^{\prime+}_2$ &$g'_0$\\
\hline\hline $t'_0$
          &$-2{l}'_0$& $2l_1'$& $2l_1^{\prime +}$ &0&$-M$&
          $ M^+$
          & 0& 0 &0\\
\hline $ t'_1$
                    &$2l_1^{\prime }$ &$4l'_2$&$-2g'_0$& $-2M$&0&
                    $-t'_0$ & 0 &
                    $-t^{\prime+}_1$     &$t'_1$\\
\hline $t^{\prime+}_1$
          & $2l_1^{\prime +}$& $-2g'_0$&$4l^{\prime+}_2$&$2M^+$&
          $t'_0$ & 0 &
          $t'_1$ & 0  &$-t^{\prime+}_1$\\
\hline\hline $l'_{0}$ &0&$2M$&$-2M^+$&0&$r{K_1^0}^+$&$-r{K}^0_1$ &
 0 &0 &0 \\
\hline $l'_{1}$ &$M$& 0 & $-t'_0$&$-r{{K}^0_1}^+$&0& $X$ & 0 &
$-l^{\prime+}_1$ & $l'_1$ \\
\hline $l^{\prime+}_{1}$ &$-M^+$& $t'_0$ & 0 &$r{K}^0_1$&$-X$& 0 &
$l'_1$ &0&
$-l^{\prime+}_1$ \\
\hline $l'_{2}$
&0&0&$-t'_1$& 0 & 0 & $-l'_1$ & 0 & $g'_0$ & $2l'_2$ \\
\hline $l^{\prime+}_{2}$ & 0 & $t^{\prime+}_1$ & 0 & 0&
$l^{\prime+}_1$ & 0 & $-g'_0$ &0&
$-2l^{\prime+}_2$\\
\hline $g'_{0}$
&0&$-t'_1$&$t^{\prime+}_1$&0&$-l'_1$&$l^{\prime+}_1$
&$-2l'_2$&$2l^{\prime+}_2$
&0\\
\hline\hline
\end{tabular}
\end{center}
 \caption{Multiplication table
for $\mathcal{A}^{\prime}(Y(1),AdS_d)$ (with an explicit form of
$F^{K}_{IJ}(o')$)} \label{table'}
\end{table}
where the symbol $'+'$ at $t^{\prime+}_1$, $l^{\prime+}_i$, $i=1,2$
means a special Hermitian conjugation which will be
specify later on, and the nonlinear part of the commutator relations
is given by the formulae
\begin{align}
&\hspace{-1ex} [{t}'_0, l'_1\} = -
r\bigl[(g'_0-\textstyle\frac{1}{2})t'_1+2t^{\prime +}_1l'_2\bigr]
= - M , &   [l^{\prime +}_1, t'_0\} = - r\bigl[t^{\prime
+}_1(g'_0-\textstyle\frac{1}{2})+2l^{\prime +}_2t'_1\bigr] = -M^+
 , \label{MM+} \\
&\hspace{-1ex} [l'_0,t'_1\}   =
2r\bigl[(g'_0-\textstyle\frac{1}{2})t'_1+2t^{\prime +}_1l'_2\bigr]
= 2 M, & [l'_0,t^{\prime +}_1\}   = -2r\bigl[t^{\prime
+}_1(g'_0-\textstyle\frac{1}{2})+2l^{\prime +}_2t'_1\Bigr] = -2M^+
, \label{2MM+}\\
&\hspace{-1ex} [l'_0,l'_1\} = 2r\bigl[(g'_0
-\textstyle\frac{1}{2})l'_1+2l^{\prime +}_1l'_2\bigr] =
r{{K}^0_1}^+, & [{l}'_0,l^{\prime +}_1\}  = - 2r\bigl[l^{\prime
+}_1(g'_0-\textstyle\frac{1}{2} )+2l^{\prime +}_2l'_1\bigr] =
-r{K}^0_1, \label{KK+1}
\end{align}
\vspace{-3ex}
\begin{equation}
 [l'_1,l^{\prime +}_1\}  = {l}'_0  +
\textstyle\frac{1}{2}r \bigl[-2g^{\prime{}2}_0+8l^{\prime
+}_2l'_2+g'_0 -3 t^{\prime +}_1t'_1\bigr]= X,
  \label{X}
\end{equation}
with a constant parameter $r$ being the square of
the inverse radius of AdS$_d$-space.

Note, first of all, that the superalgebra
$\mathcal{A}^{\prime}(Y(1),AdS_d)$ is derived from a modified
(without massive terms) HS symmetry superalgebra
$\mathcal{A}(Y(1),AdS_d)$ for half-integer totally-symmetric
spin-tensors $\Psi_{\mu_1\ldots\mu_n{}A}(x)$ (with Lorentz
$\mu_i=0,...,d-1$, $i=1,...n$ and Dirac $A=1,...,2^{[d/2]}$
indices, where  $[x]$ denotes the integer part of number $x$) in
the AdS$_d$-space, whose elements $o_I$ are explicitly determined
in a Fock space $\mathcal{H}$ with a dual basis coinciding with a
set of all $\Psi_{\mu_1\ldots\mu_n{}A}(x)$, $n\in \mathbb{N}_0$
for $\Psi_{\mu_1\mu_0{}A}$ $\equiv$ $\Psi_{A}$ \cite{adsfermBKR},
and satisfy the same multiplication table as the table for the
abstract elements $o'_I$ \cite{adsfermBKR,Resh08122329} with the
only difference in the quadratic terms in the r.h.s. of
supercommutators\footnote{To establish a correspondence for the
multiplication laws, it is sufficient to make a change of the
quantities $F^{K}_{IJ}(o')$ in (\ref{nly}) for $o_I'$ by
$\breve{F}^{K_1}_{IJ}(o)= f^{(1)K_1}_{IJ} -
(-1)^{\varepsilon_{K_1}\varepsilon_{K_2}}f^{(2)K_1K_2}_{IJ}o_{K_2}$
for $o_I$, which means that linear commutators for the latter
elements coincide with the former, whereas the non-linear
relations (\ref{MM+})--(\ref{X}) remain the same for $o_I$ under
the replacement $[M,M^+,\mathcal{K}_1, \mathcal{K}^+_1,
X-l'_0](o')$ $\to $ $-[M,M^+,\mathcal{K}_1, \mathcal{K}^+_1,
X-l_0](o)$.}.

Second,  the superalgebras $\mathcal{A}^{\prime}(Y(1),AdS_d)$,
$\mathcal{A}(Y(1),AdS_d)$ coincide and pass to the Lie superalgebra
$\mathcal{A}(Y(1),\mathbb{R}^{d-1,1})$ for a vanishing $r$.

Third, the quantity ${K}^0_1, ({{K}^0_1}^+)$ is the additive part
of $\mathcal{K}_1$ ($\mathcal{K}^+_1$), which, in its turn, is
derived as a differential consequence of the Casimir operator
$\mathcal{K}_0$ for a maximal Lie subsuperalgebra
$\mathcal{A}^{Lie}(Y(1),AdS_d)$ $\simeq osp(2|1)$  in
$\mathcal{A}^{\prime}(Y(1),AdS_d)$, generated by $t'_1, t^{\prime
+}_1, l'_2, l^{\prime +}_2,  g_0'$ by means of the element
$l^{\prime +}_1$ ($l^{\prime }_1$), and determined as follows,
\begin{equation}\label{K_0}
    \mathcal{K}_1 \equiv K_1^0 + K_1^1,\   K_1^i =
    [{K}_0^i, {l^{\prime +}_1}\} ,\
    \mathcal{K}_0 = K_0^0 + K_0^1 = \left({g'_0}^2-2g'_0-4l^{\prime +}_2l'_2\right)+
    \left(g'_0 + t^{\prime +}_1t'_1\right),
\end{equation}
where $K_0^0$ is the Casimir operator for the $so(2,1)\simeq
sp(2)$\footnote{Here we have observed the well-known
correspondence among unitary irreducible representations of
Lorentz algebra $so(1,d-1)$ subject to Young tableaux with n rows
$n\leq \left[\frac{d}{2}\right]$ to $sp(2n)$ algebra by means of
Howe duality \cite{Howe}, \cite{AlkGrTip}} subalgebra and $i=0,1$.

Fourth, there exist nonvanishing third structure functions
$F_{IJK}^{L_2L_3}(o')$ for the superalgebra
$\mathcal{A}^{\prime}(Y(1),AdS_d)$ resolving the Jacobi identities
(\ref{explJacob}) for $(I,J,K) = (4,5,6)$ generated by the triple
of the elements $l_0', l'_1, l^{\prime +}_1$; see for details
Refs.~\cite{adsfermBKR,Resh08122329,nlnsuperRM}.

Following the general method of constructing an auxiliary
representation for Lie algebras \cite{Liealgebra} and non-linear
algebras \cite{0206027}, arising for integer totally-symmetric HS
fields in the AdS$_d$-space, we may consider an extension of a
Cartan-like decomposition for the Lie superalgebra
$\mathcal{A}^{Lie}(Y(1),AdS_d) \simeq osp(2|1)$:
\begin{equation}\label{Cartansusy}
\mathcal{A}^{Lie}(Y(1),AdS_d)  = \{E^{-\alpha}\} \oplus \{H^i\}
\oplus \{E^\alpha\} \equiv \{t^{\prime +}_1;  l^{\prime +}_2\}
\oplus \{g_0'\} \oplus \{t^{\prime }_1; l'_2\},\footnotemark
\end{equation}\footnotetext{The direct sum $\{  l^{\prime +}_2\} \oplus \{g_0'\} \oplus
\{ l'_2\}$ presents a Cartan decomposition of the $so(2,1)\simeq
sp(2)$.}with the Cartan generator $g_0'$, and positive $E^\alpha$
and negative $E^{-\alpha}$ root vectors till a Cartan-like
decomposition for the non-linear superalgebra
$\mathcal{A}^{\prime}(Y(1),AdS_d)$
\begin{equation}\label{Cartannlsusy}
\mathcal{A}^{\prime}(Y(1),AdS_d)  = \{{E}^{-A}\} \oplus
\{H^{\hat{i}}\} \oplus \{E^A\} \equiv \{E^{-\alpha},
\textstyle\frac{l^{\prime +}_1}{m_1} \} \oplus \{g_0', t'_0,
l_0'\} \oplus \{E^{\alpha}, \frac{l^{\prime }_1}{m_1}\},
\end{equation}
with a constant real number $m_1 \neq 0$, introduced for
convenience, and making, from the physical viewpoint, all the
negative and positive root vectors as dimensionless quantities. In
comparison with a proper Cartan decomposition, from the
multiplication table~\ref{table'}, only the third property holds
true among the commutation relations
\begin{equation}\label{Cartancomrels}
[H^{\hat{i}},E^B\} = B(\hat{i})E^B, \qquad [E^A,E^{-A}\} = \sum
A^{\hat{i}} H^{\hat{i}}, \qquad [E^A,E^B\} = N^{A{}B} E^{A+B},
\end{equation}
which characterize a Lie algebra in a Cartan--Weyl basis. Here,
$A^{\hat{i}}$, $B(\hat{i})$ and $N^{A{}B}$ play the role of
parameters, roots and structure constants of the algebra. In spite
of this fact, the last property is still sufficient to enlarge the
method of Verma module construction \cite{Liealgebra, 0206027} to
the non-linear superalgebra under consideration.

Consider the highest-weight representation of
$\mathcal{A}^{\prime}(Y(1),AdS_d)$, with the highest-weight vector
$|0\rangle_V$ annihilated by the positive roots and being the
proper vector of the Cartan generators $H^{\hat{i}}$:
\begin{equation}\label{eigenvector}
 E^{A}|0\rangle_V = 0\,, \alpha >0\,, \qquad
 \bigl(g_0^{\prime }, t'_0, l_0'\bigr)|0\rangle_V =
 \bigl(h, \tilde{\gamma}m_0, m_0^2\bigr) |0\rangle_V\,,
\end{equation}
where $\tilde{\gamma}$ is the odd $2^{[\frac{d}{2}]}\times
2^{[\frac{d}{2}]}$ supermatrix subject to the property
$\tilde{\gamma}^2 = -\mathbf{1}$, and, due to the relation
${t'_0}^2 = - l_0'$, the proper eigenvalue for $l_0'$ is
functionally dependent from the one for $t'_0$\footnote{In the rest of
the paper, we will not specify the supermatrix structure of the
elements $o'_I$.}. Following the Poincare--Birkhoff--Witt theorem,
the basis space of this representation, called in the mathematical
literature the Verma module \cite{Dixmier}, is given by the
vectors
\begin{equation}
\label{basisV} \left|{n}_1^0, {n}_2, n_3\rangle_V \right.
 = \bigl(E^{\prime - A_{1}}\bigr){}^{n^0_1}\bigl(E^{\prime -
A_{2}}\bigr){}^{n_2}\bigl(E^{\prime -
A_{3}}\bigr){}^{n_{3}}|0\rangle_V \,,
\end{equation}
where we have fixed the ordering of the positive ``roots'' $A_{1},
A_{2}, A_{3}$ and $n_2, n_3 \in \mathbb{N}_0$, ${n}_1^0 = 0, 1$
because of the  identity: $[E^{\prime - A_{1}}, E^{\prime -
A_{1}}\}$ = $4E^{\prime - A_{2}}$ $ \Longleftrightarrow [t^{\prime
+}_1, t^{\prime +}_1\}$ = $4l^{\prime +}_2$.

Using the commutation relations of the superalgebra given by
Table~\ref{table'} and the formula for the product of graded
operators $A$, $B$, for $s = \varepsilon(B)$ and $n\geq 0$,
\begin{eqnarray}
\label{product} &&    AB^n =
\sum^{n}_{k=0}(-1)^{\varepsilon(A)\varepsilon(B)(n-k)}C^{(s)}{
}^n_k B^{n-k}\mathrm{ad}^k_B{}A\,,   \   \mathrm{ad}^k_B{}A=
[[...[A,\stackrel{ k{\,} {\rm times}}{ \overbrace{B\},...\},B}\}},
\end{eqnarray}
first obtained in \cite{ mixfermi}, we can calculate the explicit
form of the Verma module. Eq.~(\ref{product}) presents generalized
coefficients for a number of graded combinations, $C^{(s)}{
}^n_k$, that coincide with the standard ones for the bosonic
operator $B$: ${C^{(0)}{}^n_k} = C^n_k = \frac{n!}{k!(n-k)!}$.
Remind that these coefficients are defined recursively by the
relations
\begin{eqnarray}\label{combination}
&&    C^{(s)}{}^{n+1}_{k} = (-1)^{s(n+k+1)}C^{(s)}{ }^n_{k-1} +
C^{(s)}{}^n_k \,, \qquad
 n, k \geq 0\,,\\
&&  C^{(s)}{}^{n}_{0} = C^{(s)}{}^{n}_{n}=1\,, \qquad
C^{(s)}{}^{n}_{k}=0\,, \ n<k\ \mathrm{or}\ k<0
\end{eqnarray}
and possess the properties $C^{(s)}{}^n_k=C^{(s)}{}^n_{n-k}$.
Explicitly, the  values of $C^{(1)}{}^n_k$ are defined, for $n\geq
k$, by the formulae
\begin{equation}\label{expressions}
C^{(1)}{}^n_k =
\sum^{(n-k+1)}_{i_k=1}\sum^{(n-i_k-k+2)}_{i_{k-1}=1}\ldots
\sum^{(n- \sum^{k}_{j=3}i_j-1)}_{i_2=1}
\sum^{(n-\sum^{k}_{j=2}i_j)}_{i_1=1} (-1)^{k(n+1) +
\textstyle\sum\limits^{[(k+1)/2]}_{j=1}\left(i_{2j-1}+1\right)},
\end{equation}
which follow by induction. It is interesting to note that
the corresponding odd analog of the Pascal triangle has a more sparse
form as compared to the standard even Pascal triangle and is given
by Table~\ref{oPastriangle} with accuracy up to the
number $C^{(1)}{}^9_k$ of odd combinations,
\begin{table}
\begin{center}
\begin{tabular}{rrr r r rrr rrr  r r r r r r r r r  r r r r r}
         &&&&&&& && &1& & &&&&&&&&&&&\\
& & &&&& & &&1&& 1&&  &&&&&&&&&\\
& & && & && &1&& 0 &&1&&&&&&&\\
& & && && &1& &1&  &1&&1&&&&&&\\

 &&  & &&&1&  &0&&2&&0&&1 &&&&&\\
 && & & &1& &1&  &2&&2&&1&&1&&&&\\
& &&&1&  & 0&& 3 &&0&&3&&0&&1&&&\\
 &&&1&  & 1&& 3 &&3&&3&&3&&1&&1&&\\
 &&1&&0&  & 4&& 0 &&6&&0&&4&&0&&1&\\
 &1&&1&  & 4&& 4 &&6&&6&&4&&4&&1&&1\\
& & ...& & ... & &
&...& &...&  &&... &&...  &&...&&...\\
\end{tabular}
\end{center}
 \caption{Odd Pascal triangle} \label{oPastriangle}
\end{table}
where the $l$-th row is composed from the values of $C^{(1)}{}^l_0$,
$C^{(1)}{}^l_1$,..., $C^{(1)}{}^l_l$, whose sum is subject to an
easy-to-prove relation:
\begin{equation}\label{oddsum}
 \sum_{k=0}^lC^{(1)}{}^l_k =
2^{\textstyle\left[\frac{l+1}{2}\right]}\  \mathrm{for\ any}\ l
\in \mathbb{N}_0.\footnotemark
\end{equation}\footnotetext{Property (\ref{oddsum}) reflects the fact that
the fermionic numbers appear by the "square root" from the bosonic
numbers corresponding  for the standard (even) Pascal triangle:
$\sum\limits_{k=0}^lC^l_k = 2^l\  \mathrm{for\ any}\ l \in
\mathbb{N}_0.$}

For the purpose of Verma module construction, due to $n^0_1 =
0,1$ in (\ref{basisV}), (\ref{product}), it is sufficient to know
that $C^{(1)}{}^0_0 = C^{(1)}{}^1_0 = 1$ and $C^{(1)}{}^{n^0_k}_1
= n^0_k$.

Returning to the calculation of the action of $o'_I$ on the basis
vectors $\left|{n}_1^0, {n}_2, n_3\rangle_V \right.$, we, first of
all, find the action of the negative root vectors $E^{-A}$  and $g_0'$:
\begin{eqnarray}
t_1^{\prime+}\left|{n}_1^0, {n}_2, n_3\rangle_V \right. & =
&\textstyle
\left(1+\left[\frac{n_1^0+1}{2}\right]\right)\left|n_1^0+ 1{}
mod{} 2,n_2 + \left[\frac{n_1^0+1}{2}\right], n_3\rangle_V\right.,
\label{t1+V}
\\
 l_2^{\prime+}\left|{n}_1^0, {n}_2, n_3\rangle_V \right. & = &
\left|{n}_1^0, {n}_2 + 1, n_3\rangle_V
\right.,  \label{l2+V}\\
 l_1^{\prime+}\left|{n}_1^0, {n}_2, n_3\rangle_V \right. & = &
m_1\left|{n}_1^0, {n}_2, n_3+1\rangle_V \right. , \label{l1+V}
\\
 g_0'\left|{n}_1^0, {n}_2, n_3\rangle_V \right. & = & (n_1^0+
2n_2+ n_3+h)\left|{n}_1^0, {n}_2, n_3\rangle_V \right..
\label{g0V}
\end{eqnarray}
Second, the intermediate result of the action of the positive root
vectors and of the remaining Cartan generators $t_0', l_0'$ on
$\left|{n}_1^0, {n}_2, n_3\rangle_V \right.$ has the form
\begin{eqnarray}
\label{toVint} {t}_0'\left|{n}_1^0, {n}_2, n_3\rangle_V \right.
&=&
 (-1)^{n_1^0}\Bigl[ -2m_1
{n_1^0}\left|{n}_1^0-1, {n}_2, n_3+1\rangle_V \right.
+(t_1^{\prime+})^{n_1^0}
(l_2^{\prime+})^{n_2}{t}_0'|0,0,n_3\rangle_V\Bigr] ,
\\
\label{t1Vint} t_1'\left|{n}_1^0, {n}_2, n_3\rangle_V \right. &=&
(-1)^{n_1^0}\Bigl[2{n_1^0}\bigl(2{n_2}+{n_3} + h)\bigr)
\left|{n}_1^0-1, {n}_2, n_3\rangle_V \right. -
{n_2}\textstyle\Bigl(1 +
 \hspace{-0.2em}\left[\frac{n_1^0+1}{2}
\hspace{-0.2em}\right]\Bigr) \times \nonumber \\
& &  \hspace{-0.5em}\textstyle\times\left|{n}_1^0 + 1{} mod{} 2,
n_2-1 +
 \hspace{-0.2em}\left[\frac{n_1^0+1}{2}
\hspace{-0.2em}\right]
 , n_3\right\rangle_V +
(t_1^{\prime+})^{n_1^0}(l_2^{\prime+})^{n_2}t'_1\left|0,0,n_3\rangle_V
\right]\Bigr],
\\
{l}_0'\left|{n}_1^0, {n}_2, n_3\rangle_V \right. &=&
  - 2r n_1^0 \left({ n_3} + h - \textstyle\frac{1}{2} \right)
\left|{n}_1^0, {n}_2, n_3\rangle_V \right.  \nonumber \\
& & +(t_1^{\prime+})^{n_1^0}(l_2^{\prime+})^{n_2} \Bigl({l}_0'-2r
n_1^0t_1^{\prime+}t_1^{\prime}\Bigr)|0,0, n_3\rangle_V,
\label{l0Vint}
\end{eqnarray}
\begin{eqnarray}
l_1'\left|{n}_1^0, {n}_2, n_3\rangle_V \right. &=& \hspace{-0.2em}
- m_1{n_2} |n_1^0,n_2-1,n_3+1\rangle_V\hspace{-0.1em}
+\hspace{-0.1em}
(t_1^{\prime+})^{n_1^0-1}(l_2^{\prime+})^{n_2}\Bigl(\hspace{-0.2em}
t_1^{\prime+}l_1'\hspace{-0.1em}-\hspace{-0.1em}
n_1^0{t}_0^{\prime}\hspace{-0.2em}\Bigr)|0,0,n_3\rangle_V\hspace{-0.1em},
\label{l1Vint} \\
l_2'\left|{n}_1^0, {n}_2, n_3\rangle_V \right. &=&
{n_2}\bigl({n_1^0} + {n_3}+n_2+h-1\bigr)
\left|n^0_1,n_2-1,n_3\rangle_V\right.
\nonumber \\
& & +(t_1^{\prime+})^{n_1^0-1}(l_2^{\prime+})^{n_2}
\Bigl(t_1^{\prime+}l'_2 -
{n_1^0}t_1^{\prime}\Bigr)|0,0,n_3\rangle_V \label{l2Vint}.
\end{eqnarray}

Third, to complete the above calculation we need to find the
result of the action of $t_0', l_0'$ and of the positive root
vectors $E^{A}$ on the vector $|0,0,n_3\rangle_V$. To this end,
the $n$-th power of the action of operator $\mathrm{ad}_{
l^{\prime +}_1}$ on $\mathcal{K}_0$ (\ref{K_0}) denoted as
$\mathcal{K}_n \equiv \mathrm{ad}^n_{l^{\prime +}_1}\mathcal{K}_0$
yields a formula for $\ n\in \mathbb{N}$,
\begin{equation}\label{K_n}
    \mathcal{K}_n =  \left(-8rl_2^{\prime +}\right)^{\left[(n-1)/2\right]}
    \sum^{\left[(n+1)/2\right]}_{m=1}\left(\mathcal{K}_2 \delta_{n,2m}+
    \mathcal{K}_1\delta_{n,2m-1}\right),
\end{equation}
where the operators $\mathcal{K}_1, \mathcal{K}_2$ are defined by
the formulae
\begin{eqnarray}\label{K_i}
    \mathcal{K}_p & = & K_p^0 + K_p^1,\ K_p^i = \mathrm{ad}_{l^{
    \prime +}_1}
    K_{p-1}^i, p=1,2, i = 0,1, \\
    \mathcal{K}_1 & = & K_1^0 + K_1^1 =\left[4l_2^{\prime
+}l_1^{\prime}+  l_1^{\prime +}(2g'_0-1)\right]+\left[l_1^{\prime
+} - t_1^{\prime +}{t}'_0 \right],  \label{K_1} \\
\mathcal{K}_2 & = & K_2^0 + K_2^1 =\left[4l_2^{\prime
+}K_2^{\prime 0}+ 2{l_1^{\prime +}}^2\right]+rl_2^{\prime
+}\left[ 1-2 K_0^{1}\right], \label{K_2} \\
K_2^{\prime 0}  &=& \mathrm{ad}_{l^{\prime +}_1}l_1^{\prime }  =
   {l}'_0 -
\textstyle\frac{1}{2}r \left(2\mathcal{K}_0+K_0^1\right)
.\label{K_2_0}
\end{eqnarray}
Then relations (\ref{K_n})--(\ref{K_2_0}) are sufficient to define
the commutation rules for the quantities $t_0', l_0'$ and for the
positive root vectors $E^{A}$ with
$\left(\frac{l_1^{\prime+}}{m_1}\right)^{n_3}$ in the form
\begin{eqnarray}\label{t0l1+}
{t}_0' \textstyle\left(\frac{l_1^{\prime+}}{m_1} \right)^{n_3} & =
& \textstyle\left(\frac{l_1^{\prime+}}{m_1} \right)^{n_3} t_0' +
\textstyle\frac{n_3}{m_1}\textstyle\left(\frac{l_1^{\prime+}}{m_1}
\right)^{n_3-1} r t_1^{\prime +}(K_0^1 - \textstyle\frac{1}{2}) +
r t_1^{\prime +}
    \displaystyle\sum\limits^{\left[n_3/2\right]}_{m=1}
    \left(-2r l_2^{\prime
+}\right)^{m-1} \nonumber \\
& & \times \textstyle\left(\frac{l_1^{\prime+}}{m_1}
\right)^{n_3-2m-1}\textstyle\left(\frac{1}{m_1}
\right)^{2m}\left[\textstyle\frac{l_1^{\prime+}}{m_1}
C^{n_3}_{2m}K_1^1 - \textstyle\frac{2rl_2^{\prime +}}{m_1}
C^{n_1}_{2m+1}(K_0^1 - \textstyle\frac{1}{2})\right]
,  \\
t_1'\textstyle\left(\frac{l_1^{\prime+}}{m_1}\right)^{n_3} & = &
\textstyle\left(\frac{l_1^{\prime+}}{m_1}\right)^{n_3} t_1'-
\textstyle\frac{n_3}{m_1}\textstyle\left(\frac{l_1^{\prime+}}{
m_1} \right)^{n_3-1}{t}_0' -
 rt_1^{\prime +}
    \displaystyle\sum\limits^{\left[n_3/2\right]}_{m=1}
    \left(-2rl_2^{\prime
+}\right)^{m-1}\nonumber \\
 & &\times \textstyle\left(\frac{l_1^{\prime+}}{m_1}
\right)^{n_3-2m-1}\textstyle\left(\frac{1}{m_1}
\right)^{2m}\left[\textstyle\frac{l_1^{\prime+}}{m_1}
C^{n_3}_{2m}(K_0^1 - \textstyle\frac{1}{2}) + \textstyle\frac{
1}{m_1}C^{n_3}_{2m+1}K_1^1\right]
 ,\label{t1l1+} \\
 {l}_0'\textstyle\left(\frac{l_1^{\prime+}}{m_1}\right)^{n_3} &
 =&  \textstyle\left(\frac{l_1^{\prime+}}{m_1}\right)^{n_3}{l}_0'
 - r\displaystyle\sum\limits^{\left[(n_3-1)/2\right]}_{m=0}
 \left(-8rl_2^{\prime
+}\right)^{m}\textstyle\left(\frac{l_1^{\prime+}}{m_1}
\right)^{n_3-2m-2}\textstyle\left(\frac{1}{m_1}
\right)^{2m+1} \nonumber \\
&&\times  \left[\textstyle\frac{l_1^{\prime+}}{m_1}
C^{n_3}_{2m+1}\left(\mathcal{K}_1 -
\textstyle\frac{1}{4^m}K_1^1\right) +
\textstyle\frac{1}{m_1}C^{n_3}_{2m+2}\left(\mathcal{K}_2 -
\textstyle\frac{1}{4^m}K_2^1\right)\right] ,\label{l0l1+} \\
 l_1'\textstyle\left(\frac{l_1^{\prime+}}{m_1}\right)^{n_3} &
 =& \textstyle\left(\frac{l_1^{\prime+}}{m_1}\right)^{n_3}\hspace{-0.2em}l_1'+
 \textstyle\frac{n_3}{m_1}\textstyle\left(\frac{l_1^{\prime+}}{m_1
 }\right)^{n_3-1}\hspace{-0.2em}K_2^{\prime 0} - 2
r\displaystyle\sum\limits^{\left[n_3/2\right]}_{m=1}\hspace{-0.2em}\left(
-8rl_2^{\prime
+}\right)^{m-1}\hspace{-0.1em}\textstyle\left(\frac{l_1^{\prime+}}{m_1}
\right)^{n_3-2m-1}\hspace{-0.2em}\textstyle\left(\frac{1}{m_1}
\right)^{2m} \nonumber \\
&& \times  \left[\textstyle\frac{l_1^{\prime+}}{m_1}
C^{n_3}_{2m}\left(\mathcal{K}_1 -
\textstyle\frac{1}{4^m}K_1^1\right) +
\textstyle\frac{1}{m_1}C^{n_3}_{2m+1}\left(\mathcal{K}_2 -
\textstyle\frac{1}{4^m}K_2^1\right)\hspace{-0.2em}\right] ,\label{l1l1+} \\
 l_2'\textstyle\left(\frac{l_1^{\prime+}}{m_1}\right)^{n_3}
  & = & \textstyle\left(\frac{l_1^{\prime+}}{m_1}\right)^{n_3}\hspace{-0.1em}l_2'
  - \textstyle\frac{n_3}{m_1}\textstyle\left(\frac{l_1^{
  \prime+}}{m_1}
\right)^{n_3-1}\hspace{-0.2em}l_1^{\prime } - \textstyle\frac{1}{
m_1^2}C^{n_3}_{2}\textstyle\left(\frac{l_1^{\prime+}}{m_1}
\right)^{n_3-2}\hspace{-0.2em}K_2^{\prime 0} + 2
r\hspace{-0.2em}\displaystyle\sum\limits^{\left[(n_3-1)/2\right]}_{
m=1}\hspace{-0.45em}\left( -8rl_2^{\prime
+}\right)^{m-1}\hspace{-0.25em} \nonumber \\
& &\hspace{-0.1em}\times
 \textstyle\left(
 \frac{l_1^{\prime+}}{m_1}
\hspace{-0.1em}\right)^{n_3-2m-2}\hspace{-0.3em}
\textstyle\left(\hspace{-0.1em} \frac{1}{m_1}
\hspace{-0.1em}\right)^{2m+1}\hspace{-0.35em}\left[
\textstyle\frac{l_1^{\prime+}}{m_1}
C^{n_3}_{2m+1}\hspace{-0.2em}\left(\mathcal{K}_1 -
\textstyle\frac{1}{4^m}K_1^1\right)\hspace{-0.2em} +
\textstyle\frac{1}{m_1}C^{n_3}_{
2m+2}\hspace{-0.2em}\left(\mathcal{K}_2 -
\textstyle\frac{1}{4^m}K_2^1
\hspace{-0.2em}\right)\hspace{-0.2em}\right]\hspace{-0.25em},
\label{l2l1+}
\end{eqnarray}
where we have taken into account that $C^n_{n+k} = 0$ for any $n,
k \in \mathbb{N}_0$. The result of the action of operators
(\ref{K_i})--(\ref{K_2_0}) and $\mathrm{ad}p_{l^{\prime
+}_1}l_0'$, $p=1,2$ on the highest-weight vector $|0\rangle_V$ is
given by the relations
\begin{eqnarray}
  \label{K01V}\left(K_0^0, K_0^1 \right)|0\rangle_V &=& \left(h(h-2), h\right)|0\rangle_V,
  \qquad \mathcal{K}_0|0\rangle_V = h(h-1)|0\rangle_V, \\
    \left(K_1^0, K_1^1\right)|0\rangle_V &=&
  \left(m_1(2h-1), m_1\right)|0,0,1\rangle_V + \tilde{\gamma}(0, m_0)|1,0,0\rangle_V , \nonumber
  \\
\mathcal{K}_1|0\rangle_V & = & 2m_1 h|0,0,1\rangle_V +
\tilde{\gamma}m_0|1,0,0\rangle_V , \\
 \left(K_2^{\prime 0}, K_2^1\right)|0\rangle_V & = & \left(
 \bigl[{m}_0^2 -
rh\bigl(h-\textstyle\frac{1}{2}\bigr)\bigr]|0\rangle_V ,  r(1-2h)|0,1,0\rangle_V\right),  \label{K2'0} \\
K_2^0 |0\rangle_V &=& 4 \bigl({m}_0^2 -
rh\bigl(h-\textstyle\frac{1}{2}\bigr)\bigr)|0,1,0\rangle_V +
2m_1^2|0,0,2\rangle_V,
 , \\
\mathcal{K}_2 |0\rangle_V &=& 4\left({m}_0^2 -
r\bigl(h^2-\textstyle\frac{1}{4}\bigr)\right)|0,1,0\rangle_V +
2m_1^2|0,0,2\rangle_V. \label{K02V}
\end{eqnarray}
 Therefore, the result of the action of ${t}_0', t'_1, {l}'_0, l'_1, l'_2$
 on  $|0, 0, n_3\rangle_V$ has the form
 \begin{eqnarray}
\label{t0n3} {{t}}_0'|0, 0, n_3\rangle_V &=& -\frac{m_1}{2}
\displaystyle\sum\limits^{\left[n_3/2\right]}_{m=1}
    \left(\frac{-2r}{
    m_1^2}\right)^{m}
C^{n_3}_{2m}|1, m-1, n_3-2m+1\rangle_V
  \\
 &{}& +\tilde{\gamma}{m}_0\displaystyle\sum\limits^{\left[n_3/2\right]}_{m=0}
    \left(\frac{-2r}{
    m_1^2}\right)^{m}
C^{n_3}_{2m} |0,m, n_3-2m\rangle_V \nonumber \\
 &{}& -\frac{m_1}{2}\displaystyle\sum\limits^{\left[n_3/2\right]}_{m=0}
    \left(\frac{-2r}{
    m_1^2}\right)^{m+1}
C^{n_3}_{2m+1}\textstyle\left(h-\frac{1}{2}\right)|1,m,n_3-2m-1\rangle_V
,\nonumber\\
\label{t1n3} t_1'|0,0, n_3\rangle_V &=& \frac{1}{2}
    \displaystyle\sum\limits^{\left[n_3/2\right]}_{m=1}
    \left(\frac{-2r}{
    m_1^2}\right)^{m}
\left(C^{n_3}_{2m}(h-\textstyle\frac{1}{2})+
C^{n_3}_{2m+1}\right)|
 1, m-1, n_3-2m\rangle_V   \\
&{}& - \tilde{\gamma}\frac{m_0}{m_1}
\displaystyle\sum\limits^{\left[n_3/2\right]}_{m=0}
    \left(\frac{-2r}{
    m_1^2}\right)^{m}
C^{n_3}_{2m+1}|0,m, n_3-2m-1\rangle_V
, \nonumber \\
\label{l0n3} {l}_0'|0,0, n_3\rangle_V &=& {m}_0^2 |0, 0,
n_3\rangle_V  - r
    \displaystyle\sum\limits^{\left[(n_3-1)/2\right]}_{m=0}
    \left(\frac{-8r}{
    m_1^2}\right)^{m} \Bigl\{C^{n_3}_{2m+1}\Bigl[(2h-4^{-m})|0, m, n_3-2m\rangle_V   \\
& & + \tilde{\gamma} \frac{{m}_0}{m_1}(1-4^{-m})|1,
m,n_3-2m-1\rangle_V\Bigr] + C^{n_3}_{2m+2}\Bigl[\frac{1}{m_1^2}
\Bigl(4\bigl[{m}_0^2 -
r\bigl(h^2-\textstyle\frac{1}{4}\bigr)\bigr]   \nonumber \\
& {} & + 2r\textstyle\left(h-\frac{1}{2}\right)4^{-m}\Bigr)|0,
m+1, n_3-2m-2\rangle_V + 2| 0, m, n_3-2m
\rangle_V\Bigr]\Bigr\}\nonumber
, \\
\label{l1n3} l_1'|0,0, n_3\rangle_V &=& \frac{m_1}{4}
    \displaystyle\sum\limits^{\left[n_3/2\right]}_{m=1}
    \left(\frac{-8r}{
    m_1^2}\right)^{m}\left[C^{n_3}_{2m}(2h-4^{-m})+ 2 C^{n_3}_{2m+1}\right]|0,
    m-1, n_3-2m+1 \rangle_V
  \\
& &  +\frac{1}{4}
    \displaystyle\sum\limits^{\left[n_3/2\right]}_{m=0}
    \left(\frac{-8r}{
    m_1^2}\right)^{m}\Bigl[\tilde{\gamma}{m}_0C^{n_3}_{2m}(1-4^{-m})|1, m-1, n_3-2m\rangle_V
\nonumber \\
& & \hspace{-0.1em} +\hspace{-0.1em}
\frac{C^{n_3}_{2m+1}}{m_1}\left\{4\bigl[{m}_0^2 -
rh\bigl(h-\textstyle\frac{1}{2}\bigr)\bigr]\hspace{-0.1em}+\hspace{-0.1em}r(1-2h)
(1-4^{-m})\right\}| 0, m, n_3-2m-1\rangle_V \Bigr]\hspace{-0.1em}
,\nonumber
\\
l_2'|0,0, n_3\rangle_V &=& -\frac{1}{4}
    \displaystyle\sum\limits^{\left[n_3/2\right]}_{m=1}
    \left(\frac{-8r}{
    m_1^2}\right)^{m}\Bigl[C^{n_3}_{2m+1}(2h-4^{-m})+ 2 C^{n_3}_{2m+2}\Bigr]|0,
    m-1, n_3-2m     \rangle_V \label{l2n3}
  \end{eqnarray}
\vspace{-3ex}
\begin{eqnarray}
\phantom{l_2'|0,0, n_3\rangle_V}& & -  \frac{1}{4}
    \displaystyle\sum\limits^{\left[n_3/2\right]}_{m=0}
    \left(\frac{-8r}{
    m_1^2}\right)^{m}\Bigl[\tilde{\gamma}\frac{{m}_0}{m_1}C^{n_3}_{2m+1}(1-4^{-m})|1, m-1, n_3-2m-1\rangle_V
\nonumber \\
& &  +  \frac{C^{n_3}_{2m+2}}{m_1^2}\left(4\bigl[{m}_0^2 -
rh\bigl(h-\textstyle\frac{1}{2}\bigr)\bigr]+r(1-2h)
(1-4^{-m})\right)|0, m, n_3-2m-2\rangle_V \Bigr]\nonumber
 .
\end{eqnarray}
Finally, relations (\ref{K01V})--(\ref{K02V}),
(\ref{t0n3})--(\ref{l2n3}) allow one to obtain from Eqs.
(\ref{toVint})--(\ref{l2Vint}) an explicit Verma module
representation $V_{\mathcal{A}'}$ for the superalgebra
$\mathcal{A}^{\prime}(Y(1),AdS_d)$, in addition to Eqs.
(\ref{t1+V})--(\ref{g0V}):
 \begin{eqnarray}
\label{t0nV} {t}_0'\left|n^0_1,n_2, n_3\rangle_V\right. &=& -
(-1)^{n_1^0}\Bigl[2m_1n^0_1\left|n^0_1-1,n_2,
n_3+1\rangle_V\right.+\frac{m_1}{2}\textstyle\Bigl(1 +
 \hspace{-0.2em}\left[\frac{n_1^0+1}{2}
\hspace{-0.2em}\right]\Bigr)\Bigl\{
\displaystyle\sum\limits^{\left[n_3/2\right]}_{m=1}
    \hspace{-0.2em}\left(\frac{-2r}{
    m_1^2}\right)^{m}
\nonumber \\
&{}& \hspace{-0.2em}\times
C^{n_3}_{2m}\hspace{-0.2em}\textstyle\left|n^0_1+1{}mod{}2,
n_2+m-1 +
 \hspace{-0.2em}\left[\frac{n_1^0+1}{2}
\hspace{-0.2em}\right], n_3-2m+1\rangle_V\right.\hspace{-0.2em}+
\hspace{-0.2em}\displaystyle\sum\limits^{\left[n_3/2\right]}_{m=0}
    \hspace{-0.2em}\left(\frac{-2r}{
    m_1^2}\hspace{-0.1em}\right)^{m+1}
 \nonumber \\
 &{}&
 \times C^{n_3}_{2m+1}\textstyle\left(h-\frac{1}{2}\right)
\left|n^0_1+1{}mod{}2,n_2+m+\left[\frac{n_1^0+1}{2}
\hspace{-0.2em}\right], n_3-2m-1\rangle_V\right.\Bigr\}\Bigr]\nonumber \\
 &{}& +\tilde{\gamma}{m}_0\displaystyle\sum\limits^{\left[n_3/2\right]}_{m=0}
    \left(\frac{-2r}{
    m_1^2}\right)^{m}
C^{n_3}_{2m} \left|n_1^0,n_2+m,n_3-2m\rangle_V\right.,
\end{eqnarray}
\vspace{-3ex}
\begin{eqnarray}
\label{t1nV} t_1'\left|n^0_1,n_2, n_3\rangle_V\right. &=&
(-1)^{n_1^0}\Bigl[2{n_1^0}\bigl(2{n_2}+{n_3} + h)\bigr)
\left|{n}_1^0-1, {n}_2, n_3\rangle_V \right. -
{n_2}\textstyle\Bigl(1 +
 \hspace{-0.2em}\left[\frac{n_1^0+1}{2}
\hspace{-0.2em}\right]\Bigr) \times \nonumber \\
& &  \hspace{-0.2em}\textstyle\times\left|{n}_1^0 + 1{} mod{} 2,
n_2-1 +
 \hspace{-0.2em}\left[\frac{n_1^0+1}{2}
\hspace{-0.2em}\right]
 , n_3\right\rangle_V +
   \displaystyle\frac{1}{2}\textstyle\Bigl(1 +
 \hspace{-0.2em}\left[\frac{n_1^0+1}{2}
\hspace{-0.2em}\right]\Bigr)\hspace{-0.2em}
    \displaystyle\sum\limits^{\left[n_3/2\right]}_{m=1}
    \left(\frac{-2r}{
    m_1^2}\right)^{m}
\nonumber \\
& & \hspace{-0.2em} \times
\Bigl(\hspace{-0.1em}C^{n_3}_{2m}\bigl(h-\textstyle\frac{1}{2}\bigr)
+ C^{n_3}_{2m+1}\hspace{-0.1em}\Bigr)
\hspace{-0.3em}\textstyle\left|{n}_1^0 + 1{} mod{} 2, n_2+m-1 +
 \hspace{-0.2em}\left[\frac{n_1^0+1}{2}
\hspace{-0.2em}\right] , n_3-2m\rangle_V\right.\hspace{-0.2em}\Bigr]  \nonumber \\
&{}&  - \tilde{\gamma}\frac{m_0}{m_1}
\displaystyle\sum\limits^{\left[n_3/2\right]}_{m=0}
    \left(\frac{-2r}{
    m_1^2}\right)^{m}
C^{n_3}_{2m+1}\left|n_1^0,n_2+m, n_3-2m-1\rangle_V\right. ,
\end{eqnarray}
\vspace{-3ex}
\begin{eqnarray}
{l}_0'\left|n^0_1,n_2, n_3\rangle_V\right. &=&
{m}_0^2\left|n^0_1,n_2,n_3\rangle_V\right. - r
\displaystyle\sum\limits^{\left[(n_3-1)/2\right]}_{m=0}
    \left(\frac{-8r}{
    m_1^2}\right)^{m} \Bigl\{\Bigl[C^{n_3}_{2m+1}(2h-4^{-m})
    \nonumber \\
&{}&
    +2 C^{n_3}_{2m+2}\Bigr]\left|n_1^0,n_2+ m,n_3-2m\rangle_V\right.
      +(-1)^{n_1^0}C^{n_3}_{2m+1}\tilde{\gamma}\frac{{m}_0}{m_1}(
      1-4^{-m})\nonumber \\
&{} & \hspace{-0.2em}\times\textstyle\Bigl(1 +
 \hspace{-0.2em}\hspace{-0.2em}\left[\frac{n_1^0+1}{2}
\hspace{-0.2em}\right]\Bigr)\left|n_1^0+1{}mod{}2, n_2+m+
 \hspace{-0.2em}\left[\frac{n_1^0+1}{2}
\hspace{-0.2em}\right], n_3-2m-1\rangle_V\right. \hspace{-0.2em}+
\displaystyle\frac{C^{n_3}_{2m+2}}{m_1^2}  \nonumber \\
& {} & \hspace{-0.2em}\times\hspace{-0.2em}\Bigl[4\bigl[{m}_0^2 -
r\bigl(h^2-\textstyle\frac{1}{4}\bigr)\bigr]\hspace{-0.1em}+\hspace{-0.1em}
2r\bigl(h-\frac{1}{2}\bigr)4^{-m}\Bigr]\hspace{-0.2em}\left|n_1^0,
n_2+ m+1 ,n_3-2m-2\rangle_V\right. \hspace{-0.2em}\Bigr\}
 \nonumber \\
& {} & -2rn^0_1\Bigl[
\displaystyle\sum\limits^{\left[n_3/2\right]}_{m=0}
    \left(\frac{-2r}{
    m_1^2}\right)^{m}
\Bigl(C^{n_3}_{2m}\bigl(h-\textstyle\frac{1}{2}\bigr)+
C^{n_3}_{2m+1}\Bigr)\left|
n^0_1,n_2+ m, n_3-2m\rangle_V\right.  \nonumber \\
&{}& -2\tilde{\gamma}\frac{m_0}{m_1}
\displaystyle\sum\limits^{\left[n_3/2\right]}_{m=0}
    \left(\frac{-2r}{
    m_1^2}\right)^{m}
C^{n_3}_{2m+1}\textstyle\left|n_1^0-1,n_2+m+1,
n_3-2m-1\rangle_V\right.\Bigr] ,\label{l0nV}\\
l_1'\left|n^0_1,n_2, n_3\rangle_V\right. &=&  -
m_1n_2\left|n^0_1,n_2-1,n_3+1\rangle_V\right. + \frac{m_1}{4}
    \displaystyle\sum\limits^{\left[n_3/2\right]}_{m=1}
    \left(\frac{-8r}{
    m_1^2}\right)^{m} \nonumber \\
&{} & \times  \Bigl[C^{n_3}_{2m}(2h-4^{-m})+ 2
C^{n_3}_{2m+1}\Bigr]\left|n_1^0, n_2+ m-1, n_3-2m+1
\rangle_V\right.
 \nonumber \\
&{} &  +\frac{1}{4}
    \displaystyle\sum\limits^{\left[n_3/2\right]}_{m=0}
    \left(\frac{-8r}{
    m_1^2}\right)^{m}\Bigl[(-1)^{n_1^0}\tilde{\gamma}{m}_0
    \textstyle\Bigl(1 +
 \hspace{-0.2em}\left[\frac{n_1^0+1}{2}
\hspace{-0.2em}\right]\Bigr)
C^{n_3}_{2m}(1-4^{-m})\nonumber \\
& & \times \textstyle\left|n_1^0+1{}mod{}2,n_2+
m-1+\hspace{-0.2em}\left[\frac{n_1^0+1}{2}
\hspace{-0.2em}\right],n_3-2m\rangle_V\right. + C^{n_3}_{2m+1}
\frac{1}{m_1}
 \nonumber \\
 &{} & \times
\Bigl(4\bigl[{m}_0^2 -
r\bigl(h^2-\textstyle\frac{1}{4}\bigr)\bigr]+r(h-\frac{1}{2})4^{-m}
\Bigr)\left|n_1^0,  n_2+m ,n_3-2m-1\rangle_V\right. \Bigr]
\nonumber
\end{eqnarray}
\vspace{-3ex}
\begin{eqnarray}
\phantom{l_1'\left|n^0_1,n_2, n_3\rangle_V\right.}  &{}& +
n^0_1\Bigl[\frac{m_1}{2}
\displaystyle\sum\limits^{\left[n_3/2\right]}_{m=1}
    \left(\frac{-2r}{
    m_1^2}\right)^{m}
C^{n_3}_{2m}\left|n_1^0, n_2+m-1,n_3-2m+1\rangle_V\right.
\nonumber \\
&{}& +
(-1)^{n_1^0}\tilde{\gamma}{{m}_0}\displaystyle\sum\limits^{\left[n_3/2\right]}_{m=0}
    \left(\frac{-2r}{
    m_1^2}\right)^{m}
C^{n_3}_{2m} \left|n_1^0-1,n_2+ m, n_3-2m\rangle_V\right.
\nonumber \\
&{}& +
\frac{m_1}{2}\displaystyle\sum\limits^{\left[n_3/2\right]}_{m=0}
    \left(\frac{-2r}{
    m_1^2}\right)^{m+1}
C^{n_3}_{2m+1}\textstyle\bigl(h-\frac{1}{2}\bigr)\left|n_1^0,n_2+
m, n_3-2m-1\rangle_V\right.\Bigr]
,\label{l1nV} \\
l_2'\left|n_1^0,n_2, n_3\rangle_V\right. &=&
{n_2}\bigl(n_2+{n_3}+h-1\bigr)\left|n^0_1,n_2-1,n_3\rangle_V\right.
\nonumber \\
{}&{}& \hspace{-0.2em}-\frac{1}{4}\hspace{-0.2em}
    \displaystyle\sum\limits^{\left[n_3/2\right]}_{m=1}
    \hspace{-0.25em}\left(\frac{-8r}{
    m_1^2}\right)^{m}\hspace{-0.25em}\Bigl[C^{n_3}_{2m+1}(2h-4^{-m})+ 2
    C^{n_3}_{2m+2}\Bigr]   \hspace{-0.3em}\left|n_1^0,n_2+
     m-1,n_3-2m\rangle_V\right.
     \nonumber \\
& & -   \frac{1}{4}
    \displaystyle\sum\limits^{\left[n_3/2\right]}_{m=0}
    \hspace{-0.2em}\left(\frac{-8r}{
    m_1^2}\right)^{m}\Bigl[(-1)^{n_1^0}\textstyle\Bigl(1 +
 \hspace{-0.2em}\left[\frac{n_1^0+1}{2}
\hspace{-0.2em}\right]\Bigr)C^{n_3}_{2m+1}\tilde{\gamma}
    \textstyle\frac{{m}_0}{m_1}(1-4^{-m})\nonumber\\
    &&\times \left|n_1^0+1{}mod{}2,
    n_2+ m-1 +
 \hspace{-0.2em}\left[\frac{n_1^0+1}{2}
\hspace{-0.2em}\right],n_3-2m-1\rangle_V\right. \nonumber \\
& & \hspace{-0.2em}+
\frac{C^{n_3}_{2m+2}}{m_1^2}\Bigl(\hspace{-0.2em}4\bigl[{m}_0^2 -
r\bigl(h^2-\textstyle\frac{1}{4}\bigr)\bigr]+r\bigl(h-\frac{1}{2}\bigr)4^{-m}
\hspace{-0.1em}\Bigr)\hspace{-0.3em}\left| n_1^0,
n_2+m,n_3-2m-2\rangle_V\right. \hspace{-0.2em}\Bigr]
 \nonumber \\
 &{}& -\frac{{n_1^0}}{2}\Bigl[
    \displaystyle\sum\limits^{\left[n_3/2\right]}_{m=1}
    \left(\frac{-2r}{
    m_1^2}\right)^{m}
\Bigl[C^{n_3}_{2m}(h-\textstyle\frac{1}{2})+
C^{n_3}_{2m+1}\Bigr)\left|
 n_1^0, n_2+m-1,n_3-2m\rangle_V\right. \nonumber \\
&{}& +(-1)^{n_1^0}2\tilde{\gamma}\frac{m_0}{m_1}
\displaystyle\sum\limits^{\left[n_3/2\right]}_{m=0}
    \left(\frac{-2r}{
    m_1^2}\right)^{m}
C^{n_3}_{2m+1}\left|n^0_1-1,n_2+m, n_3-2m-1\rangle_V\right.\Bigr]
 \label{l2nV}.
\end{eqnarray}
The set of relations (\ref{t1+V})--(\ref{g0V}),
(\ref{t0nV})--(\ref{l2nV}) completely resolves the first problem of
the paper.

\emph{\textbf{Corollary:}} For the Lie superalgebra
$\mathcal{A}^{\prime}(Y(1),\mathbb{R}^{d-1,1})$ =
$\left.\mathcal{A}^{\prime}(Y(1),AdS_d)\right\vert_{r=0}$,
the Verma module $V_{\mathcal{A}'}$ is reduced to
$V_{\mathcal{A}'}\vert_{r=0}$, having the same dimension and given by
relations (\ref{t1+V})--(\ref{g0V}) for negative root vectors
and $g_0'$, whereas for positive root vectors and $t_0', l_0'$
it is given by the following relations as a result of
the operators' action on $\left|n_1^0,n_2,n_3\rangle_V\right.$:
 \begin{eqnarray}
\label{t0nVMin} {t}_0'\left|n^0_1,n_2, n_3\rangle_V\right. &=& -
(-1)^{n_1^0}2m_1n^0_1\left|n^0_1-1,n_2,
n_3+1\rangle_V\right.+\tilde{\gamma}{m}_0
\left|n_1^0,n_2,n_3\rangle_V\right., \\
\label{t1nVMin}
t_1'\left|n^0_1,n_2, n_3\rangle_V\right. &=&
(-1)^{n_1^0}\Bigl[2{n_1^0}\bigl(2{n_2}+{n_3} + h)\bigr)
\left|{n}_1^0-1, {n}_2, n_3\rangle_V \right.  \nonumber \\
& &  \textstyle - {n_2}\textstyle\Bigl(1 +
 \hspace{-0.2em}\left[\frac{n_1^0+1}{2}
\hspace{-0.2em}\right]\Bigr)\left|{n}_1^0 + 1{} mod{} 2, n_2-1 +
 \hspace{-0.2em}\left[\frac{n_1^0+1}{2}
\hspace{-0.2em}\right]
 , n_3\right\rangle_V \Bigr]  \nonumber \\
&{}&  - \tilde{\gamma}\frac{m_0}{m_1} {n_3}\left|n_1^0,n_2,
n_3-1\rangle_V\right. , \\
{l}_0'\left|n^0_1,n_2,
n_3\rangle_V\right. &=& {m}_0^2\left|n^0_1,n_2,n_3\rangle_V\right.
 ,\label{l0nVMin}
\\
l_1'\left|n^0_1,n_2, n_3\rangle_V\right. &=&  -
m_1n_2\left|n^0_1,n_2-1,n_3+1\rangle_V\right. +
n_3\frac{m_0^2}{m_1}
 \left|n_1^0,  n_2 ,n_3-1\rangle_V\right.
\nonumber \\
&{}& + n^0_1 (-1)^{n_1^0}\tilde{\gamma}{{m}_0} \left|n_1^0-1,n_2,
n_3\rangle_V\right.
,\label{l1nVMin} \\
l_2'\left|n_1^0,n_2, n_3\rangle_V\right. &=&
{n_2}\bigl(n_2+{n_3}+h-1\bigr)\left|n^0_1,n_2-1,n_3\rangle_V\right.\hspace{-0.2em}
 - \hspace{-0.2em}  \frac{m_0^2}{2m_1^2}
     n_3(n_3-1) \hspace{-0.1em}\left| n_1^0, n_2,n_3-2\rangle_V\right.
 \nonumber \\
 &{}& -{{n_1^0}}n_3(-1)^{n_1^0}\tilde{\gamma}\frac{m_0}{m_1}
\left|n^0_1-1,n_2, n_3-1\rangle_V\right.
 \label{l2nVMin}.
\end{eqnarray}

Let us turn to the solution of the second problem.

\subsubsection{Oscillator realization of $V_{\mathcal{A}'}$ over
the Heisenberg--Weyl superalgebra}\label{Oscilator}

To this end, following the results of \cite{Liealgebra},
initially elaborated for a simple Lie algebra
and then enlarged to a special non-linear quadratic algebra in
Ref.~\cite{0206027}, we make use of the mapping for an arbitrary
basis vector of Verma module
\begin{eqnarray}\label{map}
   & \left|{n}_1^0, {n}_2, n_3\rangle_V \right.
    \longleftrightarrow \left|{n}_1^0, {n}_2, n_3\rangle \right.
 = \bigl(f^+\bigr){}^{n^0_1}\bigl(b^{+}_2\bigr){}^{
 n_{2}}\bigl(b^{+}_1\bigr){}^{n_3}|0\rangle\,, &\\
&  \mathrm{for}\ f|0\rangle = b_1|0\rangle = b_2 |0\rangle = 0,
 \quad (\left|0, 0, 0\rangle \right. \equiv |0\rangle
 ).&
\end{eqnarray}
Here $\left|{n}_1^0, {n}_2, n_3\rangle\right.$, for ${n}_1^0 =
0,1$, $n_{2}, n_3 \in \mathbb{N}_0$ together with the vacuum
vector $|0\rangle$, are the basis vectors of a Fock space
$\mathcal{H}'$ generated by 1 pair of fermionic, $f^+, f$,
 and 2 pairs of bosonic, $b^{+}_{1}, b^+_2, b_{1}, b_2$,  creation and
 annihilation operators (whose number coincides with one of the positive root vectors $E^A$), being the basis elements of the Heisenberg--Weyl
 superalgebra $A_{1,2}$, with the standard
(only nonvanishing) commutation relations
\begin{equation}\label{commrelations}
 \{f\,, f^+\} = 1\,,\qquad [b_k, b^+_l] =
 \delta_{kl}, \ k, l = 1,2\,.
\end{equation}
Then, the generators of $V_{\mathcal{A}'}$ can be represented as
formal power series in the generators of the Heisenberg--Weyl
 superalgebra. To
realize this problem, we need the following additive
correspondence among Verma module vectors and Fock space
$\mathcal{H}'$ vectors:
\begin{eqnarray}
&&   (-1)^{n_1^0}\textstyle\Bigl(1 +
 \hspace{-0.2em}\left[\frac{n_1^0+1}{2}
\hspace{-0.2em}\right]\Bigr)
C^{n_3}_{2m}\textstyle\left|n^0_1+1{}mod{}2, n_2+m-1 +
 \hspace{-0.2em}\left[\frac{n_1^0+1}{2}
\hspace{-0.2em}\right], n_3-2m+1\rangle_V\right.
\longleftrightarrow
\nonumber\\
&& \qquad - (f^+ - 2 b_2^+f)
\frac{(b_2^+)^{m-1}(b_1)^{2m-1}}{(2m)!}
\left|{n}_1^0, {n}_2, n_3\rangle \right., \label{exmap1} \\
   && {n_1^0}
\Bigl(C^{n_3}_{2m}+ C^{n_3}_{2m+1}\Bigr)\left|
 n_1^0, n_2+m-1,n_3-2m\rangle_V\right. \longleftrightarrow
\nonumber\\
&& \qquad  f^+f \Bigl\{\frac{1}{(2m)!}+\frac{b^+_1b_1}{(2m+1)!}
\Bigr\}{(b_2^+)^{m-1}(b_1)^{2m}}
\left|{n}_1^0, {n}_2, n_3\rangle \right., \label{exmap2}\\
 && (-1)^{n_1^0} {n_1^0} C^{n_3}_{2m+1}\left|n^0_1-1,n_2+m,
n_3-2m-1\rangle_V\right. \longleftrightarrow -
f\frac{(b_2^+)^{m}(b_1)^{2m+1}}{(2m+1)!} \left|{n}_1^0, {n}_2,
n_3\rangle \right..\label{exmap3}
\end{eqnarray}
The above relations are sufficient to realize the form of the
elements $o'_I \in \mathcal{A}'(Y(1),AdS_d)$ satisfying the
multiplication table~\ref{table'} as formal power series
$o'_I(b_i,b_i^+,f,f^+), i=1,2$ with respect to the degrees of
non-supercommuting generating elements, as follows (see
Ref.\cite{adsfermBKR}):
\begin{align}
\label{t1'+} &t_1^{\prime+} = f^+ +2b_2^+f, &&l_1^{\prime+}=m_1
b_1^+,
\\
\label{g0'} & g_0'= b_1^+b_1+2b_2^+b_2+f^+f+h, && l_2^{\prime+}
=b_2^+,
\end{align}
\vspace{-2ex}
\begin{eqnarray}
t_0' &=& 2m_1b_1^+f -\frac{m_1}{2}(f^+{-}2b_2^+f)\,b_1^+
    \sum_{k=1}^{\infty}
    \left(\frac{-2r}{m_1^2}\right)^{k}
    \frac{(b_2^+)^{k-1}b_1^{2k}}{(2k)!}
+\tilde{\gamma}m_0\sum_{k=0}^{\infty}
    \left(\frac{-2r}{m_1^2}\right)^{k}
    \frac{(b_2^+)^{k}b_1^{2k}}{(2k)!}
\nonumber
\\
&&{} +\frac{r(h-{\textstyle\frac{1}{2}})}{m_1}(f^+{-}2b_2^+f )
    \sum_{k=0}^{\infty}\left(\frac{-2r}{m_1^2}\right)^k
    \frac{(b_2^+)^{k}\;b_1^{2k+1}}{(2k+1)!}
, \label{t0'}
\\
t_1' &=& -2 g_0'f -(f^+{-}2b_2^+f)b_2
+\frac{1}{2}(h-{\textstyle\frac{1}{2}})(f^+{-}2b_2^+f)
    \sum_{k=1}^{\infty}
    \left(\frac{-2r}{m_1^2}\right)^{k}
    \frac{(b_2^+)^{k-1}b_1^{2k}}{(2k)!}
\nonumber
\\
&&{} +\frac{1}{2}(f^+{-}2b_2^+f)\;b_1^+
    \sum_{k=1}^{\infty}
    \left(\frac{-2r}{m_1^2}\right)^{k}
    \frac{(b_2^+)^{k-1}\;b_1^{2k+1}}{(2k+1)!}
\nonumber
\\
&&{} -\frac{\tilde{\gamma}m_0}{m_1}\;\sum_{k=0}^{\infty}
    \left(\frac{-2r}{m_1^2}\right)^{k}
    \frac{(b_2^+)^{k}\; b_1^{2k+1}}{(2k+1)!}
\label{t1'} ,
\end{eqnarray}
\vspace{-2ex}
\begin{eqnarray}
l_0' &=& m_0^2 -r\frac{\tilde{\gamma}m_0}{m_1}\;(f^+{-}2b_2^+f)
    \sum_{k=1}^{\infty}
    \left(\frac{-8r}{m_1^2}\right)^{k}
    \frac{(b_2^+)^k\;b_1^{2k+1}}{(2k+1)!}\;(1-4^{-k})
\nonumber
\\
&&{} -rb_1^+
    \sum_{k=0}^{\infty}
    \left(\frac{-8r}{m_1^2}\right)^{k}
    \frac{(b_2^+)^k\;b_1^{2k+1}}{(2k+1)!}\;(2h-4^{-k})
+4r\; \frac{\tilde{\gamma}m_0}{m_1}\;f\;
    \sum_{k=0}^{\infty}
    \left(\frac{-2r}{m_1^2}\right)^{k}
    \frac{(b_2^+)^{k+1}\;b_1^{2k+1}}{(2k+1)!}\nonumber
    \\
    &&{} + {r\left(h-\frac{1}{2}\right)}
    \sum_{k=0}^{\infty}
    \left(\frac{-2r}{m_1^2}\right)^{k+1}
    \frac{(b_2^+)^{k+1}\;b_1^{2k+2}}{(2k+2)!}
-2r\;(b_1^+)^2\sum_{k=0}^{\infty}
    \left(\frac{-8r}{m_1^2}\right)^k
    \frac{(b_2^+)^k\;b_1^{2k+2}}{(2k+2)!}
 \nonumber
\\
 &&{} -2rf^+f\;
    \sum_{k=0}^{\infty}
    \left(\frac{-2r}{m_1^2}\right)^{k}
    \left\{\frac{(h-\textstyle\frac{1}{2})}{(2k)!}
    +\frac{b_1^+b_1}{(2k+1)!}\right\}(b_2^+)^{k}b_1^{2k}
\nonumber
\\
&&{} +   \frac{m_0^2-r(h^2-\frac{1}{4})}{2}
    \sum_{k=0}^{\infty}
    \left(\frac{-8r}{m_1^2}\right)^{k+1}
    \frac{(b_2^+)^{k+1}\;b_1^{2k+2}}{(2k+2)!}
, \label{l0'}
\\
l_1' &=& -m_1b_1^+b_2 +\frac{m_1}{4}\;b_1^+\;
    \sum_{k=1}^{\infty}
    \left(\frac{-8r}{m_1^2}\right)^{k}
    \left\{\frac{2h-4^{-k}}{(2k)!}
   +\frac{2b_1^+b_1}{(2k+1)!}\right\}(b_2^+)^{k-1}b_1^{2k}
+ \nonumber
\\
&&{} +\frac{\tilde{\gamma}m_0}{4}\;(f^+{-}2b_2^+f)
    \sum_{k=1}^{\infty}
    \left(\frac{-8r}{m_1^2}\right)^{k}
    \frac{(b_2^+)^{k-1}\;b_1^{2k}}{(2k)!}\;(1-4^{-k})
\nonumber
\\
&&{} + \frac{r(h-\frac{1}{2})}{2m_1}
    \sum_{k=0}^{\infty}
    \left(\frac{-2r}{m_1^2}\right)^{k}
    \frac{(b_2^+)^{k}\;b_1^{2k+1}}{(2k+1)!}
+\frac{m_1}{2}\;b_1^+f^+f\;
     \sum_{k=1}^{\infty}
    \left(\frac{-2r}{m_1^2}\right)^{k}
    \frac{(b_2^+)^{k-1}b_1^{2k}}{(2k)!}
\nonumber
\\
&&{} -\frac{r(h-{\textstyle\frac{1}{2}})}{m_1}f^+f
    \sum_{k=0}^{\infty}
    \left(\frac{-2r}{m_1^2}\right)^{k}
    \frac{(b_2^+)^{k}b_1^{2k+1}}{(2k+1)!}
-\tilde{\gamma}m_0f\sum_{k=0}^{\infty}
    \left(\frac{-2r}{m_1^2}\right)^{k}
    \frac{(b_2^+)^{k}b_1^{2k}}{(2k)!}
\nonumber
\\
&&{} +\frac{m_0^2-r(h^2-\frac{1}{4})}{m_1}
    \sum_{k=0}^{\infty}
    \left(\frac{-8r}{m_1^2}\right)^{k}
    \frac{(b_2^+)^{k}\;b_1^{2k+1}}{(2k+1)!}
, \label{l1'} \\
l_2' &=& g_0'b_2 - b_2^+b_2^2
-\frac{m_0^2-r(h^2-\frac{1}{4})}{m_1^2}\;
    \sum_{k=0}^{\infty}
    \left(\frac{-8r}{m_1^2}\right)^{k}
    \frac{(b_2^+)^k\;b_1^{2k+2}}{(2k+2)!}
\nonumber
\\
&&{} -\frac{r(h-\frac{1}{2})}{2m_1^2}\;
    \sum_{k=0}^{\infty}
    \left(\frac{-2r}{m_1^2}\right)^{k}
    \frac{(b_2^+)^k\;b_1^{2k+2}}{(2k+2)!}
+\frac{\tilde{\gamma}m_0}{m_1}\;f\;\sum_{k=0}^{\infty}
    \left(\frac{-2r}{m_1^2}\right)^{k}
    \frac{(b_2^+)^{k}b_1^{2k+1}}{(2k+1)!}
\nonumber
\\
&&{} -  \frac{\tilde{\gamma}m_0}{4m_1}\;(f^+{-}2b_2^+f)
    \sum_{k=1}^{\infty}
    \left(\frac{-8r}{m_1^2}\right)^{k}
    \frac{(b_2^+)^{k-1}b_1^{2k+1}}{(2k+1)!}\;(1- 4^{-k})
\nonumber
\\
&&{} -\frac{1}{4}\;b_1^+\;
    \sum_{k=1}^{\infty}
    \left(\frac{-8r}{m_1^2}\right)^{k}
    \left\{\frac{2h-4^{-k}}{(2k+1)!}
    +\frac{2b_1^+b_1}{(2k+2)!}
    \right\}(b_2^+)^{k-1}b_1^{2k+1}
\nonumber
\\
&&{} - \frac{1}{2}\;f^+f\; \sum_{k=1}^{\infty}
    \left(\frac{-2r}{m_1^2}\right)^{k}
    \left\{\frac{h-\textstyle\frac{1}{2}}{(2k)!}
  + \frac{b_1^+b_1}{(2k+1)!}\right\}
    (b_2^+)^{k-1}b_1^{2k}
 \label{l2'}.
\end{eqnarray}
The infinite sums in these expressions are simple in view
of their acting on an arbitrary vector $\left|{n}_1^0, {n}_2,
n_3\rangle \right. \in \mathcal{H}'$. For instance, the second
sums in (\ref{t0'}) and (\ref{l0'}) may be written with the help of
the formal variable $x =
\bigl((2rb_2^+b_1^2)/m_1^2\bigr)^{\frac{1}{2}}$ as follows:
\begin{eqnarray}
  \tilde{\gamma}m_0\sum_{k=0}^{\infty}
    \left(\frac{-2r}{m_1^2}\right)^{k}
    \frac{(b_2^+)^{k}b_1^{2k}}{(2k)!}
 &=&  \tilde{\gamma}m_0\sum_{k=0}^{\infty}
    \left({-1}\right)^{k}
    \frac{x^{2k}}{(2k)!} = \tilde{\gamma}m_0 \cos x,
\end{eqnarray}
\begin{eqnarray}
 -rb_1^+
    \sum_{k=0}^{\infty}
    \left(\frac{-8r}{m_1^2}\right)^{k}
    \frac{(b_2^+)^k\;b_1^{2k+1}}{(2k+1)!}\;(2h-4^{-k}) &=&  {m_1b_1^+}\sqrt{\frac{
    r}{2b_2^+}}\Bigl\{\sin x - h \sin 2x\Bigr\},
\end{eqnarray}
so that the other sums in (\ref{t0'})--(\ref{l2'}) can be
rewritten as combinations of $\sin x$, $\sin 2x$, $\cos x$, $\cos
2x$. However, the representation for $o_I'$ as a formal series
power is preferably applicable to specific calculations.

The set of relations (\ref{t1'+})--(\ref{l2'}) completely resolves
the second problem of the paper.

It is suitable to note that the above Fock space realization of
the superalgebra $\mathcal{A}^{\prime}(Y(1),AdS_d)$ does not
preserve the property of the closedness of
$\mathcal{A}^{\prime}(Y(1),AdS_d)$ with respect to the standard
Hermitian conjugation,
\begin{align}
& (l_0')^+\neq l_0', && (l_i')^+\neq l_i^{\prime+}, &&
(t_0')^+\neq t_0' && (t_1')^+\neq t_1^{\prime+},
\end{align}
if one should use the standard rules \cite{Liesusy,
adsfermBKR} of Hermitian conjugation for
$b^{+}_{i}, b_{i}, f^+, f$:  $(b_{i})^+=b^{+}_{i}, (f)^+ = f^+$
and for $(\tilde{\gamma})^+ = - \tilde{\gamma}$. Therefore, to provide
the closedness of $\mathcal{A}^{\prime}(Y(1),AdS_d)$ we need to change
the standard Euclidian scalar product in the Fock space $\mathcal{H}'$,
which is expressed by an appearance of the operator $K$, whose form is
completely determined by equations which express a new
Hermitian conjugation property (see Refs. \cite{Liealgebra,
adsfermBKR}) for $o'_I(b_i,b_i^+,f,f^+)$,
\begin{align}
K \bigl(E^{- A}\bigr)^{+}= E^{A} K, && K \bigl(E^{ A}\bigr)^{+}=
E^{-A} K,  && K \bigl(H^{\hat{i}}\bigr)^+ = H^{\hat{i}}K.
\end{align}
These relations allow one to determine an operator $K$ being
Hermitian with respect to the usual scalar product, as follows:
\begin{eqnarray}
\label{explicit K} K'=Z^+Z, \qquad
Z=\sum_{({n}_2,n_3)=({0},0)}^{\infty}\sum_{{n}^0_1=0}^1
\left|{n}^0_1,{n}_2,n_3\rangle_V\right.\frac{1}{{n}_{2}!n_3!}
\langle 0|b_{1}^{n_3}b_{2}^{n_2}f^{n_1^0}.
\end{eqnarray}

\vspace{1ex} \emph{\textbf{Corollary:}} For the Lie superalgebra
$\mathcal{A}^{\prime}(Y(1),\mathbb{R}^{d-1,1})$ =
$\left.\mathcal{A}^{\prime}(Y(1),AdS_d)\right\vert_{r=0}$ the
oscillator realization of the Verma module
$V_{\mathcal{A}'}\vert_{r=0}$ given by the relations
(\ref{t1+V})--(\ref{g0V}), (\ref{t0nVMin})--(\ref{l2nVMin}) is
reduced to the polynomial realization with the same relations
(\ref{t1'+}), (\ref{g0'}) for the negative root vectors and
$g_0'$, whereas for the positive root vectors and $t_0', l_0'$ we
have
 \begin{align}
&t_0' = 2m_1b_1^+f +\tilde{\gamma}m_0, \label{t0'Min} && t_1' = -2
g_0'f -(f^+{-}2b_2^+f)b_2
 -\frac{\tilde{\gamma}m_0}{m_1} b_1\\
&l_0' = m_0^2 , \label{l0'Min} && l_1' = -m_1b_1^+b_2
-\tilde{\gamma}m_0f + \frac{m_0^2}{m_1}
    b_1, \\
&l_2' = g_0'b_2 - b_2^+b_2^2 -\frac{m_0^2}{2m_1^2}\;
    b_1^2
+\frac{\tilde{\gamma}m_0}{m_1}\;f\;b_1 . &&
 \label{l2'Min}
\end{align}
The above result for the oscillator realization of the
superalgebra $\mathcal{A}^{\prime}(Y(1),\mathbb{R}^{d-1,1})$ and,
in particular, for $osp(2|1)$ subsuperalgebra, differs from the
analogous result, given in Ref. \cite{Liesusy}.

\subsection{Formalized representation of superalgebra
$\mathcal{A}'(Y(1),AdS_d)$ (explicit formal setting of the
problem)}\label{A1AdS_d}

The Verma module $V_{\mathcal{A}'}$ for the superalgebra
$\mathcal{A}'(Y(1),AdS_d$ and its realization in terms of a formal
power series in the degrees of the elements of the
Heisenberg--Weyl superalgebra $A_{1,2}$, obtained in Sect.
\ref{VermaSUSY}, \ref{Oscilator}, require (as mentioned in
Introduction), for the correctness and reliability of the final
expressions (\ref{t1'+})--(\ref{l2'}) for $o'_I(b_i,b_i^+,f,f^+)$,
to verify the fact that they ($o'_I(b_i,b_i^+,f,f^+)$) indeed
satisfy the multiplication table~\ref{table'}. This problem is
extremely laborious as a purely mathematical process. Indeed, the
only powerful means on this way may be the method of mathematical
induction with the parameter $q$ in $r^q$, ($q\in \mathbb{N}_0$),
due to the necessity of double sum calculations arising in
supercommutators  $[o'_I(b_i,b_i^+,f,f^+),
o'_J(b_i,b_i^+,f,f^+)\}$. The problem becomes practically
unsolvable in a reasonable time by hands in view of a polynomial
(of the fourth degree) growth of the number of calculation
operations $\mathcal{N}$ [related to ones of independent
supercommutators
\begin{equation}\label{numindscom}
\mathcal{N}=\textstyle(1+k^2+\frac{5}{2}k)(3+2k^2+5k)=
4(1+k^2+\frac{5}{2}k)^2 -
(1+k^2+\frac{5}{2}k)[2(1+k^2+\frac{5}{2}k)-1],
\end{equation}
i.e. the entries of upper triangular superantisymmetric matrix in
the table~(\ref{table'}),] as one turns to the superalgebra
$\mathcal{A}^{\prime}(Y(k),AdS_d)$ with the growth of the number
of rows $k$ in the corresponding Young tableaux used for
half-integer HS fields\footnote{For  non-linear algebra
$\mathcal{A}^{\prime}_b(Y(k),AdS_d)$ corresponding to integer spin
tensors the numbers of its elements and independent commutators
are equal respectively to $(1+2k^2+3k)$ and $\mathcal{N}_b =
k\bigl(k+\frac{3}{2}\bigr)(k+1)(2k+1)$.}. So, the number of
$\mathcal{N} = 45, 210, 1170, ... $ for $k=1,2,3,...$ in
superalgebra $\mathcal{A}^{\prime}(Y(k),AdS_d)$.

Therefore, a solution of the third problem consists in a
reformulation of the representation (\ref{t1'+})--(\ref{l2'}) for
$\mathcal{A}^{\prime}(Y(1),AdS_d)$ in the formalized problem
setting applicable to the development of a computer realization
of a verification of the multiplication table~\ref{table'}
within the symbolic computation approach. In doing so, we have to take
into account a necessity to use only the restricted induction
principle with respect to the degrees of inverse square AdS$_d$-radius
$r$: $r^q$, $q = 0, 1, 2, ..., l$, because of the impossibility
of an immediate application of its mathematical analogue,
in view of the finiteness of an actual volume of memory elements.

\vspace{1ex} \noindent \emph{\textbf{The formal setting
of the algorithm (FSA)}} includes the following steps:
\begin{enumerate}
    \item computation for $\mathcal{A}^{\prime}(Y(1),AdS_d)$ of the
    products in the left-hand side, $\bigl(o'_I o'_J - (-1)^{\varepsilon_I\varepsilon_J}
    o'_Jo'_I\bigr)$ $\equiv$  $P_{IJ}^l(b_i,b_i^+$, $f, f^+)$, of supercommutators
    to be verified, $[o'_I o'_J\}$ $\equiv$  $P_{IJ}^r(b_i,b_i^+,f,f^+)$, $I,J=1,...,9$, $i=1,2$, given
    by Table~\ref{table'}, as polynomials with respect to
    non-supercom\-mu\-ting elements $b_i,b_i^+,f,f^+$ with a fixed
    maximal degree $q$ in $r$,  $r^q$, $q=0,1,...,q_0$, $q_0\in
    \mathbb{N}$, which we denote, for the leading monomials of $P_{IJ}^{l(r)}$, by
    $\mathrm{Im}_r(P_{IJ}^l) = q$,
    $\mathrm{Im}_r(P_{IJ}^r) = q$;
       \item  rearrangement of the product $P_{IJ}^l(b_i,b_i^+$, $f, f^+)$ to
       a \emph{regular
       monomial ordering}, based on an introduction of
       a \emph{monomial ordering} $\prec$ on the universal enveloping algebra
       for the Heisenberg--Weyl superalgebra $A_{1,2}$: $U(A_{1,2})$, being in
       one-to-one correspondence with the \emph{total ordering} $\prec$ on $\mathbb{N}_0\times
       \mathbb{Z}_2\times \mathbb{N}_0^2
       \times \mathbb{Z}^2\times \mathbb{N}_0$ $ \simeq \mathbb{N}_0^4\times \mathbb{Z}_2^2$,
       because of the set of monomials $\bigl\{
       \bigl((b_2^+)^{k_4}, (f^+)^{l_2}, (b_1^+)^{k_3},
       b_2^{k_2},f^{l_1},b_1^{k_1}\bigr)\bigr\}$ that forms
       a Poincare--Birkhoff--Witt (PBW) basis in
       $U(A_{1,2})$, is in bijection with $\mathbb{N}_0^4\times \mathbb{Z}_2^2$:
        \begin{equation}\label{bijection}
\bigl((b_2^+)^{k_4}, (f^+)^{l_2}, (b_1^+)^{k_3},
       b_2^{k_2},f^{l_1},b_1^{k_1}\bigr) \leftrightarrow  ({k_4}, {l_2}, {k_3},
       {k_2}, {l_1}, {k_1}), k_1,...,k_4 \in \mathbb{N}_0; l_1,l_2 \in
       \mathbb{Z}_2;
\end{equation}
    \item comparison (or calculation of the difference) of $P_{IJ}^{l}$ and
    $P_{IJ}^{r}$ for each
    fixed $q$:\\  $\mathrm{Im}_r(P_{IJ}^l) = \mathrm{Im}_r(P_{IJ}^r) = q$;
    for $q=0,1,...,q_0$.
\end{enumerate}

To solve the problem of the first item, we need to take into
account that the supercommutator $[o'_I, o'_J\}$ is an
anticommutator,
\begin{equation}\label{anticomm}
[o'_I, o'_J\} = o'_I o'_J + o'_J o'_I,\  \mathrm{iff\ } \
\varepsilon_I = \varepsilon_J = 1 \ \Leftrightarrow I,J \in
\{1,2,3\},
\end{equation}
and a commutator,
\begin{equation}\label{comm}
[o'_I, o'_J\} = o'_I o'_J - o'_J o'_I,\  \mathrm{iff\ } \
\bigl(\varepsilon_I = 0 \ \mathrm{or}\ \varepsilon_J = 0\bigr) \
\Leftrightarrow ((I \in \{4,..,9\})\ \mathrm{or}\ (J \in
\{4,..,9\})).
\end{equation}

A treatment of the second item is based on a list of properties
for the following primary elements, which do not have an internal
structure, for the purpose of the third and fourth problems:
\begin{itemize}
    \item the
quantities $m_0, m_1, r, h$ in (\ref{t1'+})--(\ref{l2'}) are
constant even elements commuting with all the others
quantities;
    \item $b_i^+, f^+, b_i, f$ are non-supercommuting
    (generating for $o'_I$) elements which satisfy properties
    (\ref{commrelations}) and additionally the following ones:
    \begin{equation}\label{nilpotf}
    [b_i,b_j]=[b_i^+,b_j^+]=[b_i,f]= [b^+_i,f]= [b^+_i,f^+]=
    [b_i,f^+]=0,\ \ f^2=(f^+)^2=0.
\end{equation}
    \item $\tilde{\gamma}$ is an odd constant quantity
    (whose matrix nature we will ignore) obeying the properties:
\begin{equation}\label{gammat}
\tilde{\gamma}^2=-1,\quad \tilde{\gamma}a=-a\tilde{\gamma},\
\mathrm{for}\ a\in\{f,f^+\},\quad
\tilde{\gamma}b=b\tilde{\gamma},\ \mathrm{for}\ b\in\{b_i,b^+_i\}.
\end{equation}
\end{itemize}
As to the bijection (\ref{bijection}) among $U(A_{1,2})$ and
$\mathbb{N}_0^4\times \mathbb{Z}_2^2$, for instance, the monomial
$\frac{1}{5!}m_1^{-2}b_2^+f^+b_1^+fb_1^3$ may be represented as
follows:
\begin{equation}\label{composition}
\frac{1}{5!}m_1^{-2}b_2^+f^+b_1^+fb_1^3  \mapsto
\frac{1}{5!}m_1^{-2}(1,1,1,0,1,3).
\end{equation}
The above list is sufficient to determine the following
easy-to-obtain formula necessary to rearrange the products
of two arbitrary monomials $a_1(b_i^+,f^+,f,b_i),
a_2(b_i^+,f^+,f,b_i)$, written in the regular monomial ordering,
for $\mathrm{Im}_r(a_1)= m \leq q_0$, $\mathrm{Im}_r(a_2)= m' \leq
q_0$, which compose arbitrary polynomials, including the
elements of $\mathcal{A}^{\prime}(Y(1),AdS_d)$, restricted by the
condition $\mathrm{Im}_r(o'_I)=q$:
\begin{eqnarray}\label{fproda2a1}
a_2\cdot a_1 &=&
r^{m'}(b_2^+)^{k'_4}(f^+)^{l'}(b_1^+)^{k'_3}b_2^{k'_2}f^{k'}b_1^{k'_1}\cdot
r^m(b_2^+)^{k_4}(f^+)^{l}(b_1^+)^{k_3}b_2^{k_2}f^kb_1^{k_1}\nonumber
\\
&=&r^{m'+m} \sum^{k'_1}_{n= \max(0,k'_1-k_3)}\sum^{k'_2}_{n'=
\max(0,k'_2-k_4)} \frac{k'_1 !}{n!(k'_1 - n)!}\frac{k_3 !}{(k_3 -
(k'_1 -
  n))!} \nonumber \\
& &\times
  \frac{k'_2 !}{n'!(k'_2 - n')!}\frac{k_4 !}{(k_4 - (k'_2 -
  n'))!}(b_2^+)^{k'_4+ k_4 -k'_2 +n'}f^{+l'}(b_1^+)^{k'_3+k_3 -k'_1
  +n}b_2^{n'+k_2} \nonumber \\
& & \times
\Bigl[f^{k'}\delta_{l0}\delta_{k'1}+f^{+l}\delta_{k'0}\delta_{l1}
  +(1-f^+f)\delta_{l1}\delta_{k'1}\Bigr] f^kb_1^{n+k_1}.
\end{eqnarray}
For $m+m'> q_0$, we must set $a_2\cdot a_1 = 0$.\footnote{Formula
(\ref{fproda2a1}) can be easily rewritten in terms of the product
of  integer-valued vectors
$({k'_4},l'_2,{k'_3},{k'_2},l'_1,{k'_1})$,
$({k_4},l_2,{k_3},{k_2},l_1,{k_1})$ which is naturally determined
due to a bijection (\ref{bijection}) of the PWB basis with
$\mathbb{N}^4 \times \mathbb{Z}^2_2$.} As a result, the product of
2 monomials (PBW basis elements) modulo the coefficient $r^m
(r^{m'})$ is expressed through a polynomial composed again from
PBW basis elements.

At last, because of the necessity to verify the multiplication
table~\ref{table'} with accuracy up to $r^q$, $q=0,1,...,q_0$ we
need the following relation:
\begin{eqnarray}\label{ruleorder}
[o'_I, o'_J\}_{k} & = & \sum^{k}_{l=0}[({o'_I})_{k-l},
({o'_J})_l\}, \ \mathrm{for} \ \mathrm{all}\ k=0,1,...q_0,\texttt{ where} \nonumber \\
 ({o'_I})_{k-l}
&=& r^{k-l}A_{k-l}, \qquad [o'_I, o'_J\}_{k}  = r^{k}B_k, \texttt{
without summation on }k, l,
\end{eqnarray}
for some completely definite quantities $A_{k-l}, B_{k}$ defined
by table~\ref{table'} and relations (\ref{t1'+})--(\ref{l2'}).

The solution of the third item of FSA is rather technical and
consists in a simultaneous visual presentation in a dialog box
of the left- and right-hand sides (or their difference) of the
verified supercommutator with a required accuracy in $r^q$.

\section{Programming realization}\label{Programmreal}

In this section, we consider the concept of programming
realization for the above-mentioned {\bf formal setting of the
algorithm}. To this end, we introduce data structures which
realize the elements of the superalgebra
$\mathcal{A}'(Y(1),AdS_d)$ and operations among them within the
\emph{object-oriented paradigm}.

\subsection{Concept and properties of the program}\label{Concept}

Starting from the purpose of automatic verification
mentioned in FSA and given in terms of algebraic
quantities, we shall realize it as a program
with the help of computer algebra methods.

As mentioned in Introduction, even in the case of Lie algebras and
superalgebras we need to use symbolic computational approach to
treat these algebraic structures in the case of their realization
as polynomials of finite order over a corresponding
Heisenberg--Weyl algebra and superalgebra, whose elements are
regarded as symbols within a programming realization. Another
point concerns the peculiarities of our programming comparison
with the module \emph{Plural}, being the most developed one in the
case of treatment of left ideals and modules over a given
non-commutative $G$\emph{-algebra}. The main peculiarities are:

\noindent 1) the treatment, on equal footing, of non-commuting
$b_i, b_i^+$ and not-anticommuting $f, f^+$ symbols of a given
Heisenberg--Weyl superalgebra (which is absent in \emph{Plural});

\noindent 2) the use of a  different realization of basic
programming procedures within the object-oriented paradigm being
the basis of the program language $C\#$.

To create our program, we simulate a superalgebra (so-called
\emph{basic model of the superalgebra}) to be applicable to the
treatment of an arbitrary non-linear associative superalgebra with
respect to the standard multiplication ``$\cdot$''. Second, we
introduce a \emph{model of polynomial superalgebra} as a special
enlargement of the basic model, taking into account the internal
structure of a concrete polynomial superalgebra, i.e. the number
of non-supercommuting basis elements of given Heisenberg--Weyl
superalgebra,  the number and polynomial structure of  basis
elements  of given superalgebra, explicit form of its
multiplication table. Third, we realize, on the basis of a
\emph{model of polynomial superalgebra}, a calculation of
to-be-verified left- and right-hand sides of commutators from
Table~\ref{table'}, and then make a comparison with a given
accuracy.

In realizing the program, we start from the requirement of its
universality. This means that the program must promote
a resolution of not only a concrete polynomial superalgebra but
also symbolic computations of arbitrary polynomials
constructed from non-supercommuting elements.

Despite the fact that the basic purpose of our program is
an automation of verification procedures, it should
be noted that completely automatic analytic calculations
pose a complicated problem. Therefore, the main task to
be solved becomes a minimization of routine work being
potentially subjected to human error. A significant issue
is a flexibility of an output of program data
for its subsequent treatment, either by a specialist or
by another program of automatic calculations. In the first
case, the data at the final stage of program work, as well as
on each stage throughout checking, must have a visual
representation in an appropriate form. In the second case,
the data have to be presented in a form available for
analysis of another program.

The main window of the program is divided into three sections,
as demonstrated by Figure~\ref{figure1} in
Section~\ref{Programm}. At the top of the window, there are
control elements which permit one to choose the left $o'_I$ and right
$o'_J$ arguments of a verified supercommutator [$o'_I, o'_J\}$.
At the bottom, there are two panels for visual means of
representation of an explicitly calculated product
$P_{IJ}^l(b_i,b_i^+$, $f, f^+)$ on the left and the (supposed)
form of the right-hand side of the supercommutator
$P_{IJ}^r(b_i,b_i^+$, $f, f^+)$ on the right panel obtained in
correspondence with the data of multiplication Table~\ref{table'}.
The graphical presentation of formulae is made with help of
the component \emph{WebBrowser}.

The program creates a specially marked HTML-document, which
illustrates the current results of calculations. The next (general
for the majority of programming products) property is processing
speed. For all of the required operations for the superalgebra
$\mathcal{A}'(Y(1),AdS_d)$ under consideration, the program
produces the result in just several minutes, which completely
satisfies requirements for its application. Indeed, even in the
case of a large size of input data, a launch of the program for
given supercommutator has a unique character. The possibility of
further optimization and improvement of the program's processing
speed will be described in Section~\ref{ConclusionP}.

\subsection{Data structures and methods}\label{DSMT}

Here, we shall introduce the notion of a two-level model and
consider in detail the methods of its treatment.

\subsubsection{Basic model of a superalgebra}\label{BMS}

Let us simulate the object of a superalgebra as
applied to the treatment of an arbitrary
(in the algebraic sense) non-linear associative
superalgebra with respect to the usual
multiplication "$\cdot$".

The model presents a realization of elements $a_1,\ldots ,  a_n$,
$n\in \mathbb{N}$ of an arbitrary $\mathbb{K}$-superalgebra with
additive and multiplicative composition laws, such that all
possible results of these operations over $a_1,\ldots ,  a_n$ are
elements of the same superalgebra
\begin{equation}\label{lincomb}
 \sum_{l_1,...,l_p}C_{l_1{}l_2{}...l_p}(a_{k_1})^{l_1}\cdot(a_{k_2})^{l_2}\ldots
 \cdot(a_{k_p})^{l_p}, \  , C_{l_1{}l_2{}...l_p}\in
 \mathbb{K},
\end{equation}
obtained  in an arbitrary order for $p$ able to satisfy the
inequality, $p\geq n$. For instance, among such elements may be
the monomial
\begin{equation}\label{exmonomial}
\frac{5}{m_1}(h-1) a_n(a_2)^2(a_3)^2a_5 (a_1)^3a_3(a_2)^3,
\end{equation}
where $m_1, h$ are some constants like those in
Eqs.~(\ref{t1'+})--(\ref{l2'}) and we will later omit the sign of
multiplication ``$\cdot$''.

\begin{figure}
\begin{center}
\includegraphics*[width=0.9\columnwidth]{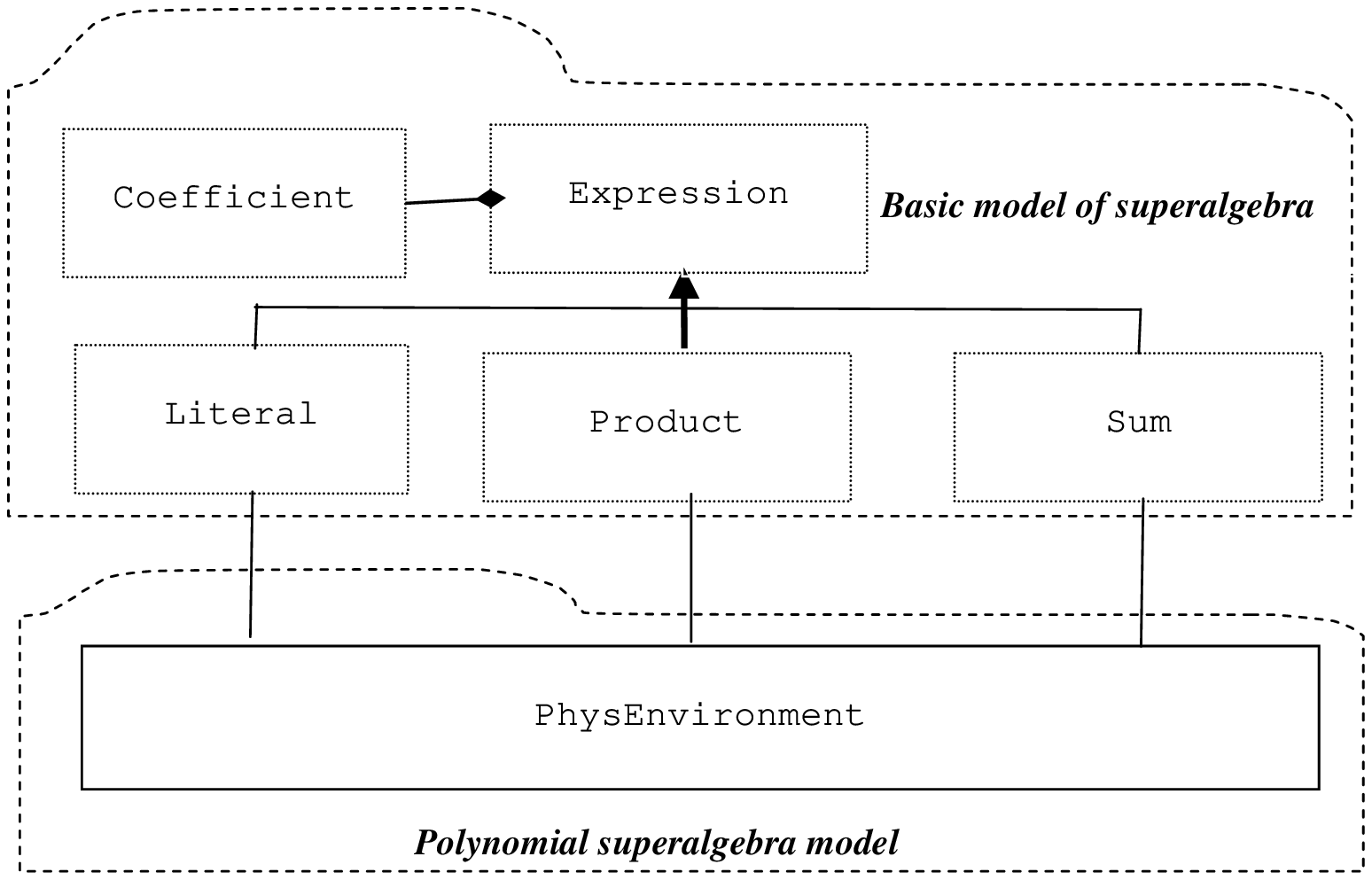}
\end{center}
\caption{Diagram of interrelations of the Classes}
\label{diarram1}
\end{figure}

At this level, the basic program data are subdivided into two types to be
 treated differently. First of them is formed by \emph{numeric coefficients}
 from the field $\mathbb{K}$ and second represent
 \emph{quantities} being the elements of a superalgebra (for instance,
 non-commuting elements of a Heisenberg--Weyl superalgebra), which
 differ from the first type by non-permutability with respect to
 the usual multiplication. They are realized within the program
 by the Classes \verb"Coefficient" and \verb"Literal". It should be noted that
 the Class \verb"Literal" is a descendant of an abstract data type
 (Class)
 \verb"Expression", which we introduce as a basic data type for
 the \emph{basic model of a superalgebra}. Each instance
 (copy) of the Class
  \verb"Expression" represents an \textbf{expression} which combines
  elements of a superalgebra, first, by means of
  summation ``$+$'' and multiplication ``$*$'', second, with the help of
  brackets of different level of multiplicity, and possesing
   a numerical coefficient from the Class
  \verb"Coefficient". The expression itself can be an element
of the  Class  \verb"Literal" representing either a product as an
element of the Class \verb"Product" or a sum as an element of the
Class \verb"Sum". Interrelations among the Classes may be
characterized by the following diagram of Classes given by
Fig.~\ref{diarram1}.

\subsubsection{Model of a polynomial superalgebra $\mathcal{A}'(Y(1),AdS_d)$}
\label{CMS}

In the last version of the system, the \emph{polynomial
superalgebra model} is realized by means of the unique Class
\verb"PhysEnvironment", which reproduces all the peculiarities of
the superalgebra $\mathcal{A}'(Y(1),AdS_d)$ not realized in the
basic model. Among them, one can select the following points:
\begin{enumerate}
    \item the set of generating elements of the Heisenberg--Weyl
    superalgebra $A_{1,2}$: $f,f^+,b_i,b_i^+$, $i=1,2$;
    \item the order of their sequence (normal ordering) determined in
    Eqs. (\ref{composition}) in
    composing the elements $o'_I$ of the superalgebra
    $\mathcal{A}'(Y(1),AdS_d)$;
    \item their ($o'_I$) explicit forms as polynomials $o'_I\bigl(
    b_2^+, f^+, b_1^+, b_2, f, b_1\bigr)$ over generating elements of
    $A_{1,2}$;
    \item calculation of the products of these polynomials with their
    normal ordering.
\end{enumerate}

\subsection{C\# Realization}\label{Csharp}

The program is realized in the computer language $C\#$ and
provides, as mentioned in Section~\ref{Concept}, a graphical
interface of calculations for specialists in
algebra. At present, it is possible to run the program using
\verb".NET Framework v.2.0" or \verb"Mono v.2.4".

In addition to the conceptual description of the objects on the
first and second levels of representation of the two-level model
for the superalgebra $\mathcal{A}'(Y(1),AdS_d)$ mentioned
respectively in Sections~\ref{BMS}, \ref{CMS}, let us consider in
some detail a realization in $C\#$ of properties and methods
(procedures) for the treatment of instances of corresponding
Classes.

The abstract Class \verb"Expression"
\begin{eqnarray}
 && \verb" public abstract class Expression"\hspace{20em}
 \label{Expression}
\end{eqnarray}
describes an \textbf{expression} as an element of the \emph{basic
model of superalgebra} and has the following important public
fields
\begin{eqnarray}
 && \verb"public Coefficient Coefficient;"\hspace{20em}\nonumber\\
 && \verb"public int Power;"\hspace{15em}\label{fieldsExpression}
\end{eqnarray}
the first of which is responsible for a numeric
\textbf{coefficient} [which can be written in the form
$\frac{d_1^{k_1}\ldots d_m^{k_m}}{c_1^{l_1}\ldots c_p^{l_p}}$,
with $k_1,...,k_m, l_1,...,l_p \in \mathbb{N}_0$ and commuting
quantities $c_1, ..., c_p, d_1,...,d_m$ being natural numbers or
special symbols like $h, m_1, m_0$ in
Eqs.(\ref{t1'+})--(\ref{l2'})] considered as an element of the
Class \verb"Coefficient". To complete the description, we only
note some interesting methods used for the treatment of
\emph{expressions} from the Class \verb"Expression":
\begin{eqnarray}
 && \verb"public abstract Expression Simplify();"\hspace{10em}
\nonumber\\
   && \verb"public abstract bool IsSimple();" \hspace{10em}\nonumber\\
    && \verb"public abstract bool SimilarTo(Expression expression);"
    \hspace{8em}
    \label{SimilatTo}
\end{eqnarray}
which results, respectively, in the returning of a new instance
from the Class \verb"Expression", equivalent (from the algebraic
viewpoint) to the previous one but having a simpler structure
which consists in an opening of algebraic brackets and in
concatenation of homogeneous objects into a unique object (such as
the sum of sums from the Class \verb"Sum" and the product of
products from the Class \verb"Product"). Simultaneously, in the
procedure \verb"IsSimple()" one realizes a verification of the
fact if it is necessary to simplify the expression and if it is
similar to another expression with respect to multiplication
``$*$''.

Omitting a description of some technical methods inherent in the
instance of the Class \verb"Coefficient", we pay attention to the
public fields
\begin{eqnarray}
 &&  \verb"public List<CoefficientItem> Numerator;" \hspace{15em}
    \nonumber\\
  && \verb"public List<CoefficientItem> Denumerator;" \hspace{15em}
  \label{Denumerator}
\end{eqnarray}
which serve for the above-mentioned representation of coefficients
as rational fractions with positive power exponents
$\frac{d_1^{k_1}\ldots d_m^{k_m}}{c_1^{l_1}\ldots c_p^{l_p}}$ by
analogy with a graphical representation of fractions in the
mathematical formulation of the problem. As an analog of the
procedure \verb"Simplify" for \verb"Expression" here appears the
method \verb"Normalize":
\begin{eqnarray}
 && \verb"public virtual void Normalize();"\hspace{20em}
  \label{Normalize}
\end{eqnarray}
which changes the visual program structure of the object
transforming it into a mathematically equivalent instance.

To determine a separate numeric coefficient of the expression,
we have introduced the Class \verb"CoefficientItem":
\begin{eqnarray}
 && \verb"public int Power;"\hspace{15em} \nonumber\\
&& \verb"public bool SimilarTo(CoefficientItem"\
\verb"Coefficient);"\hspace{10em}\label{Power}
\end{eqnarray}
characterized by the field \verb"Power" responsible
for the degree of a single multiplier in any of the coefficients.
The procedure \verb"SimilarTo" realizes a search for similar
co-multipliers with respect to multiplication.

The Class \verb"Literal" contains information on the representation
of an element of some superalgebra as a record
similar to Eq. (\ref{exmonomial}) with a field for a numeric
coefficient and other fields for symbols [at this stage
without the property of commutation as in
Eqs. (\ref{commrelations}), (\ref{nilpotf})]. Each instance from
\verb"Literal" contains a corresponding representation for
upper and lower indices, as in Eqs.
(\ref{anticomm})--(\ref{composition}):
\begin{eqnarray}
 && \verb"protected string _subIndex;"
   \hspace{15em}\nonumber\\
&&  \verb"protected string _supIndex;"\hspace{23em}
\label{subindex}
\end{eqnarray}
whereas the methods of their treatment coincide significantly with
those from the Class \verb"Coef"\-\verb"ficient" with some
specifics; for example, the method
\begin{eqnarray}
 && \verb"public override bool SimilarTo(Expression"\
 \verb"expression);" \hspace{9em}\label{Similar}
\end{eqnarray}
seeks for the same literals which differ modulo their
mathematical powers (superscripts).

In turn, the Class \verb"Product" representing the product of some
expressions is important on the second level of our two-level
program model because the product of normally ordered
polynomials in the powers of $b_i^+,f^+,b_i,f$ will determine the
element of Poincare--Birkhoff--Witt basis (\ref{basisV}) in the
oscillator representation (\ref{map}), having, after
a simplification (method \verb"Simplify"), the form of
a monomial as in Eq. (\ref{composition}). Co-multipliers of some
\emph{product} are contained as a list in the case of
\emph{expressions}:
\begin{eqnarray}
 && \verb"public List<Expression> this[int Index];"
   \hspace{16em} \nonumber\\
   && \verb"public int Length;" \hspace{15em}
\label{listprod}
\end{eqnarray}
that permit one to keep some complicated algebraic structures in
the product. Among various methods, there are some methods
inherited from the class \verb"Expression" which allow one to
concatenate in a \emph{product}  an  \emph{expression} in the case
of its multiplication by the product from the right:
\begin{eqnarray}
 &&\hspace{-1em} \verb"public static Product operator *(Product"\
\verb"left, Expression  right)"\hspace{2em} \label{prodexpres}
\end{eqnarray}
Notice that the most significant methods for \verb"Product"
are the following:
\begin{eqnarray}
 && \verb"public override bool IsSimple();"
   \hspace{17em} \nonumber\\
   && \verb"public override Expression Simplify();" \hspace{17em}
\label{Simplprod}
\end{eqnarray}
which permit one, respectively, to define a so-called \emph{simple}
product of the literals, i.e., without nested brackets, and to open
brackets with a simultaneous assignment of co-multipliers of nested
products to \emph{simple} products.

In comparison with the Class \verb"Product", the interface and methods
of treatment of instances of the Class \verb"Sum" are quite simple
and follow from the fact that they represent descendants (as well as
those of \verb"Product") of the Class \verb"Expression".
In particular, some of the methods for \verb"Sum",
\begin{eqnarray}
 && \verb"public static Sum operator +(Sum left,"\
 \verb"Expression"\
 \verb"right);"\hspace{8em}
\label{SumExpr+}\\
&& \verb"public static Sum operator *(Sum left,"\
 \verb"Sum right);"\hspace{9em} \label{SumProd}
\end{eqnarray}
determine, respectively, the rules of summation from the right of
any instance from the Class \verb"Sum" with an arbitrary
\emph{expression}  and  states  that the multiplication of sums is
the sum of the products of its summands, whereas the
\emph{coefficient} of a product is the product of
\emph{coefficients} of co-multipliers.

 Properly a \emph{model of polynomial superalgebra } for the
superalgebra $\mathcal{A}'(Y(1),AdS_d)$ as the second level of the
program model data is realized by means of the class
\verb"PhysEnvironment":
\begin{eqnarray}
&&\verb"public class PhysEnvironment"\hspace{23em}
\label{PhysEnvir}
\end{eqnarray}
whose instances are given later on with the help of a description of
string constants
\begin{eqnarray}
 && \verb"public const string QuantitySymbols = ""\Gamma\verb"bf"";\nonumber\\
&& \verb"public const string OperationalSymbols = ""\verb"tlg"";
\hspace{13em}
 \label{operatSymbols}
\end{eqnarray}
which are necessary to describe both the basis elements
$b_i,b_i^+,f,f^+$ of the superalgebra $A_{1,2}$, together with odd
quantities $\Gamma \equiv \tilde{\gamma}$, and the elements $o'_I$
of the superalgebra $\mathcal{A}'(Y(1),AdS_d)$, written here
without primes:
\begin{equation}\label{corresp1}
    \bigl[t \,, l\,,g\bigr] \longrightarrow \bigl[(t_0,t_1,t_1^+) \,,
    (l_0,l_i,l_{i}^+)\,,g_0\bigr].
\end{equation}
Especially important is the globally defined
integer-valued variable \verb"PowerLimit":
\begin{equation}
\label{r_restr} \verb"public static int PowerLimit = 1;"
\hspace{16em}
\end{equation}
which determines a restriction on the exponent in the power
$r^{q_0}$ for  elements of  $\mathcal{A}'(Y(1),AdS_d)$ as
polynomials $o'_I(b_i^+,f^+,f,b_i)$ in the powers of $r$,
$o'_I(b_i^+,f^+,f,b_i) = \sum_{k\geq 0} r^ko^{\prime
k}_I(b_i^+,f^+,f,b_i)$, for their products in the supercommutator
$([o'_I,o'_J\} = P_{IJ}^l)$ with $\mathrm{Im}_r(P_{IJ}^l)\leq
q_0$, in order to verify the validity of Table~\ref{table'} with a
given accuracy in the powers of $r$.

From the methods of treatment of instances from the class
\verb"PhysEnvironment", we consider only those which directly
determine the solution of the problem within its formal setting in
Section~\ref{A1AdS_d} and have an algebraic sense of the literals
"$(b_i^+,f^+,f,b_i)$". So the procedures
\begin{eqnarray}
 && \verb"static public bool"\
\verb"IsVanishing(Literal quantity);"\hspace{7em}\nonumber\\
&&  \verb"static public bool"\ \verb"IsCommuting(Literal left,"\
\verb"Literal right);"\hspace{7em} \label{IsCommuting}
\end{eqnarray}
realize, respectively, a verification of the nilpotency condition in
Eqs.~(\ref{nilpotf}) for $f, f^+$, and verify if two given
instances from the Class \verb"Literal" commute with each
other in correspondence with Eqs. (\ref{nilpotf}),(\ref{gammat})
in FSA. The method \verb"Commute":
\begin{eqnarray}
 && \verb"static public Expression Commute(Literal"\
 \verb"left, Literal right);" \hspace{6em}\label{Commute}
\end{eqnarray}
is a procedure of ordering of symbolic co-multipliers in a product
up to its right ordering given as in Eq.~(\ref{fproda2a1}). Given
this, if in the ordering process there are non-commuting
quantities (which is verified by the procedure
\verb"IsCommuting"), then one realizes a transformation of these
quantities according to Eqs. (\ref{commrelations}),
(\ref{nilpotf}), (\ref{gammat}).

A proper ordering of the product of an arbitrary monomials
$a_1, a_2$ is given, according to Eq. (\ref{fproda2a1}), by
means of the method
 \begin{eqnarray}
 && \verb"static public Expression SortMonomial(Product"\
 \verb"product);"\hspace{10em} \label{SortMonomial}
\end{eqnarray}
The procedure (\ref{SortMonomial}) represents the one of the basic
methods at the second level of the program model data. Let us
consider an \emph{algorithm of its work} in details.
\begin{enumerate}
    \item Check whether a given product of monomials to be an
    (incorrectly ordered) monomial with the only product of
    literals constructed from the quantities
  $\Gamma,b_i^+,f^+,b_i,f$
     (\verb"QuantitySymbols").
    \item Prepare a variable \verb"result" for the expected result
    of the algorithm.
    \item Realize the cycle over all the quantities
    $\Gamma,b_i^+,f^+,b_i,f$ that enter into the product
    \begin{description}
        \item[a)] if there is no quantity in the product then
        returns the zero;
        \item[b)] if the quantity is $\Gamma$ (i.e. $\tilde{\gamma}$)
        then we apply the rule given in Eqs. (\ref{gammat});
        \item[c)] put all other quantities into the list
        \verb"_quantities".
    \end{description}
    \item Initialize an instance of the auxiliary class
    \verb"QuantityComparer" which has a correct ordering of
    a sequence of quantities $b_2^+,f^+,b_1^+,b_2,f,b_1$.
    \item Initialize by $\mathbf{1}$ the integer-valued variable
    \verb"checkedCount" which keeps a number of quantities checked on the condition of correct ordering.
    \item cycle over the number of ordered quantities\footnote{It is worth
    noting that this cycle is similar, modulo non-supercommutativity  of
    the quantities, to the method of \emph{bubble sort},
    however, instead of a one-dimensional array (to be analogous to a monomial) we have here the
    another data structure with varying number of such "arrays" (to be similar to a polynomial).}: \begin{description}
        \item[a)] Compare the last ordered quantity with one not yet
        verified. \begin{description}
            \item[1)] If the quantities are in the wrong order, we
            check commutation properties;\begin{description}
                \item[a.] if they commute, then:
                \begin{description}
                    \item[1.] we change them by
                the places in the list \verb"_quantities" (right quantity swap
                to the left)
                    \item[2.] Now, we need to make a next checking
                    with the preceding ordered quantity. To this end,
                    we reduce \verb"checkedCount" on $\mathbf{1}$ and continue
                    the basic cycle.
                \end{description}
                \item[b.] Else, it is necessary to apply one from
                the  relations:
                (\ref{commrelations}),(\ref{nilpotf}),(\ref{gammat}),
                (\ref{fproda2a1})\begin{description}
                    \item[1.] In the product \verb"result" puts all
                    numbered by counter \verb"checkedCount"
                    correct ordered quantities.
                    \item[2.] Multiply \verb"result" by the result
                    of transformation of non-commuting quantities
                    by known rules with use of the method \verb"Commute()"
                    \item[3.] Multiply \verb"result" by all other
                    yet unchecked quantities and return  its value.
                \end{description}
            \end{description}
            \item[2)] If the quantities are in correct order,
            augment the counter \verb"checkedCount" by $\mathbf{1}$.
          \end{description}
        \end{description}
    \item If the above cycle $\mathbf{6.}$ finishes successfully, it means
    that the initial monomial is completely ordered and we return the
    product of the quantities in the sequence of its appearance to
    the list \verb"_quantities" $\Box$.
\end{enumerate}
Thus, the method \verb"SortMonomial" returns a correctly ordered
monomial, if all the elements of the initial monomial commute with
each other as in:
\begin{equation}\label{example1}
(f^+)^{l}(b_2^+)^{k_4-k_4'}(b_1^+)^{k_3-k_3'}(b_2^+)^{k_4'}b_2^{
k_2}f^k(b_1^+)^{k_3'}b_1^{k_1}, k_3'\leq k_3, k_4'\leq k_4,
\end{equation}
 or
if they have already been in the right order as in:
\begin{equation}\label{example2}
\Gamma(b_2^+)^{k_4}(f^+)^{l} (b_1^+)^{k_3}b_2^{k_2}f^kb_1^{k_1}.
\end{equation}
In other cases, it will return the result of the transformation of
the product of the quantities $\Gamma, b_i^+, f^+, f,b_i$ with
respect to known supercommutation relations, so that in a result
of a multiple application of the above algorithm one guarantees a
transformation of the initial product into a polynomial with
correctly ordered monomials.

To generate the elements $o'_I$ (\ref{t1'+})-(\ref{l2'}) of the
superalgebra $\mathcal{A}'(Y(1),AdS_d)$, polynomials
$P^l_{IJ}(b_i^+, f^+, f,b_i)$ and polynomials $P^r_{IJ}(b_i^+,
f^+, f,b_i)=[o'_I,o'_J\}$ from the cells of the multiplication
table~\ref{table'} whose formal power series are restricted by the
value of \verb"PowerLimit" (\ref{r_restr}), we have elaborated
corresponding methods:
\begin{eqnarray}
&& \verb"static public Expression GetRelation(Literal"\
\verb"operationalQuantity)";\hspace{4em}
\nonumber\\
 && \verb"static public Expression  GetPredictedOperationalProduct(int"\nonumber\\
&& \qquad\qquad\qquad\qquad \hspace{-1ex} \verb"leftOperatorIndex, int rightOperatorIndex)";\nonumber\\
&&\verb"static public Expression GetPredictedFormula(int"\
\verb"formulaIndex)". \label{Getpredicdtedformula}
\end{eqnarray}
In the two last procedures, the arguments are the values of
indices of the co-multipliers $o'_I$, $o'_J$: $I,J=1,...,9$
determined in Eq. (\ref{numoperators}) and the number of the
formula  in Table~\ref{table'} which contains the result of
calculation of $[o'_I, o'_J\}$.

The following high-level methods
 \begin{eqnarray}
 && \verb"static public Expression SolveRelations(Expression"\
\verb"expression);"\hspace{5em}\nonumber\\
&& \verb"static public Sum SortPolynomial(Sum"\
\verb"polynomial);"\label{SortPolyn}\nonumber\\
 && \verb"static public Expression"\
 \verb"SolveOperationalProduct(Literal"\ \nonumber\\
&& \qquad\qquad\qquad\qquad \hspace{-1ex}\verb"leftOperation,"
\verb"Literal rightOperation)" \label{SortPolyn}
\end{eqnarray}
result in a program realization of the formal setting for the
algorithm stated in the Section~\ref{A1AdS_d}. Indeed, the first
method  waits to get as an argument the commutator (\ref{comm}) or
anticommutator (\ref{anticomm}) of $o'_I, o'_J$ and returns the
result of total transformation of a given supercommutator $[o'_I,
o'_J\}$. The second one orders the monomials in a given
polynomials on a basis of the above-described method
\verb"SortMonomial" and realizes a restriction for the value of
the exponent $q$ in the powers of $r^q$ for a given polynomial. At
last, the third method in (\ref{SortPolyn}) serves for the
multiplication  of the operator $o'_I, o'_J$ of the given
superalgebra $\mathcal{A}'(Y(1),AdS_d)$, while taking into account
the restriction on $q$ in the product of two correctly ordered
polynomials in correspondence with Eqs. (\ref{ruleorder}).

We have thus described the program's realization in the language
$C\#$ for basic data structures and methods of their treatment
as a two-level program model which solves the \emph{formal
setting of the algorithm}.

\section{Application to verification of the algebraic properties
of $V_{\mathcal{A}'}$ }\label{Programm}

\begin{figure}
\begin{center}
\includegraphics*[width=0.9\columnwidth]{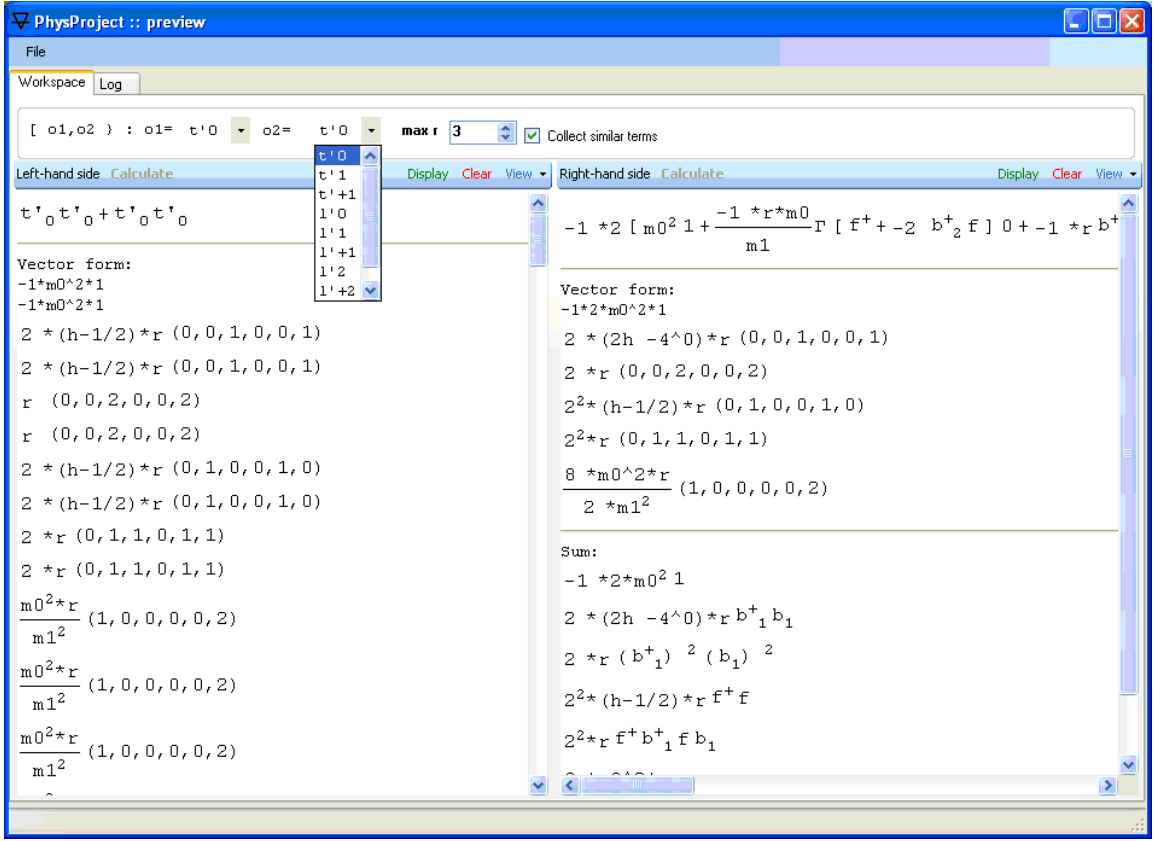}
\end{center}
\caption{Main window of the application PhysProject}
\label{figure1}
\end{figure}

We now list the subproblems solved by the program
\emph{PhysProject} within a solution of the basic problem of
verification of the multiplication table~\ref{table'} for the
elements $o'_I(b_i^+,f^+,b_i,f)$ (\ref{t1'+})--(\ref{l2'}) of the
non-linear superalgebra $\mathcal{A}'(Y(1),AdS_d)$ constructed
from the Verma module $V_{\mathcal{A}'}$.
\begin{enumerate}\item[1.] The program simulates, on the second
level of the program's data model,
    explicit forms of operators
    $o'_I(b_i^+,f^+,b_i,f)$ with a given accuracy in the degrees of
    the inverse
    squared  radius of the AdS$_d$-space, $r$, as
    polynomials in the powers of the generating elements of the
    Heisenberg--Weyl superalgebra $A_{1,2}$.
\end{enumerate}
    This is easily shown, as illustrated by Figure~\ref{figure1},
as one chooses as the second $o'_2$ (first $o'_1$) multiplier in
the verified supercommutator the operator $t_0'$, and sets the
value of the maximal degree in $r$ in the corresponding window
with counter \emph{\textbf{max} r}.

Evidently, by adding new rules of generation some
operators like $o'_I(b_i^+,f^+,b_i,f)$ [possibly with other
generating elements] we are able to adapt the program
\emph{PhysProject} to other non-linear superalgebras.

On the level of realization, the formulae are given in a form
sufficiently close to that used in its initial mathematical
description such as \emph{single-line form}, when all the monomials
in the formula are written in one line (as at the top of the right
panel on Figure~\ref{figure1}), in the \emph{vector form} like
to Eq. (\ref{composition}) (as in the middle of the right panel
and in the left panel on the figure), in the \emph{symbolic form}
with one monomial in the line (as at the bottom of the right panel
of the figure).
\begin{enumerate}
    \item[2.] The program produces an automatic simplification of
    the explicit form of elements $o'_I$ with a given accuracy in the powers of
    $r$ and calculates the product of any two elements $o'_I, o'_J$,
representing the result in a normal ordering form, when all of the
generating elements of the
    Heisenberg--Weyl superalgebra in the product are written in such a way that
    the creation operators ($b_2^+, f^+, b_1^+$) follow in their
    writing before the annihilation operators ($b_2, f, b_1$)\footnote{See
    the right panel on Figure~\ref{figure1}, where the expression for the
    operator $-2l'_0$ is written as one checks the validity
    of the supercommutator $[t'_0, t'_0\}$ up to the 3rd power in
    $r$.}.
    \item[3.] The problem of collecting similar summands has
    not yet been completely solved at present due to
    non-mathematical types of numeric coefficients; however, the program permits one
    to reduce the opposite summands. To this end, one uses the toggle "Collect similar
    items" on the main window of \emph{PhysProject}.
    \item[4.] The program produces a visual representation of
    the obtained results after some choice of the maximal degree on $r$
    and elements $o'_1,
    o'_2$, whose supercommutator should be verified. Then the
    result of the \emph{left-hand-side} window reserved for the
    $P^l_{IJ}$ polynomial (left-hand side value of $[o'_I,o'_J\}$)
    in question and the one in the \emph{right-hand-side}
    window for the $P^r_{IJ}$ (right-hand side value of $[o'_I,o'_J\}$)
    with accuracy up to value in "\textbf{max r}" are computed
    after calling of the corresponding
    procedures by means of the buttons "\emph{Calculate}". As a
    basic way of output of the results, we use a \emph{symbolic
    form} which may be chosen from the above-described 3 options
    in the list "\emph{View}" in the top from the right of both the windows.
\end{enumerate}

As a final result of the work of the program, we obtain by a
direct comparison of verified expressions from the left- and
right-hand sides of the main window that  all the relations from
the multiplication table~\ref{table'} for the superalgebra
$\mathcal{A}'(Y(1),AdS_d)$ with the elements given by Eqs.
(\ref{t1'+})--(\ref{l2'}) are valid with accuracy up to the fourth
power in $r$. Because of a cyclic manner of definition the
corresponding polynomials $o'_I(b_i^+,f^+,b_i,f)$ (i.e. following
to restricted induction principle), using the program, whose
maximal degree is restricted by the value of $q$ in $r^q$, we may
argue that the multiplication law for the elements of a
superalgebra under consideration is true.

\section{Conclusions and Perspectives to $\mathcal{A}(Y(k),AdS_d), k>1$}
  \label{ConclusionP}

In the present work, we have solved a number of problems, which do
not seem closely related at first glance, both in a purely
algebraic direction and within the area of symbolic computations,
which at the same time are related to each other from High Energy
Physics considerations.

Initially, we have realized the Verma module $V_{\mathcal{A}'}$
construction \cite{Dixmier}, applied here to the non-linear
superalgebra $\mathcal{A}'(Y(1),AdS_d)$ introduced in Ref.
\cite{adsfermBKR} and serving a Lagrangian formulation for massive
higher-spin spin-tensors in AdS$_d$-spaces as elements of
irreducible AdS-group representation space, characterized by an
arbitrary Young tableaux with one row. Within a system of
definitions introduced here in order to classify a set of
non-linear Lie-type superalgebra, the superalgebra
$\mathcal{A}'(Y(1),AdS_d)$ appears by a polynomial superalgebra of
order $2$. The construction of Verma module is based on a
generalized Cartan procedure following from the fact that negative
root vectors ($t^{\prime +}_1, l^{\prime +}_2$) from the maximal
Lie subsuperalgebra $osp(2|1)$ in $\mathcal{A}'(Y(1),AdS_d)$ are
enlarged by an operator $l^{\prime +}_1$ determining the nonlinear
part of the latter superalgebra. Formulae
(\ref{t1+V})--(\ref{g0V}), (\ref{t0nV})--(\ref{l2nV}) completely
solve the problem of Verma module construction. In the case of the
Lie superalgebra $\mathcal{A}'(Y(1),\mathbb{R}^{d-1,1})$, we have
obtained a new, in comparison with that of Ref. \cite{Liesusy}
(where it was used the Verma module for $osp(2|1)$ then enlarged
to one for $\mathcal{A}'(Y(1),\mathbb{R}^{d-1,1})$  by means of
dimensional reduction from $\mathbb{R}^{d-1,2}$ to
$\mathbb{R}^{d-1,1}$), realization of Verma module, given by Eqs.
(\ref{t1+V})--(\ref{g0V}), (\ref{t0nVMin})--(\ref{l2nVMin}). Note
that during the investigation of this problem we have obtained
some interesting results, such as \emph{Odd Pascal triangle},
given by Table~\ref{oPastriangle}, and determined by the same
rules as its standard even analog but with the help of a number of
odd-valued combinations (\ref{expressions}).

We have realized the Verma
    module $V_{\mathcal{A}'}$ in terms of a formal power series in
    the degrees of non-supercommuting generating elements
    $b_i,b_i^+,f,f^+, i=1,2$
    of a Heisenberg--Weyl superalgebra $A_{1,2}$,
    whose number coincides with those
    of negative and positive
    root vectors in a Cartan-like decomposition
    for the superalgebra $\mathcal{A}^{\prime}(Y(1),AdS_d)$. This
    problem is completely described by the formulae
    (\ref{t1'+})--(\ref{l2'}). The corresponding oscillator
    realization for the Lie superalgebra
    $\mathcal{A}'(Y(1),\mathbb{R}^{d-1,1})$ has a
    polynomial form given by Eqs. (\ref{t1'+}), (\ref{g0'}), (\ref{t0'Min})--(\ref{l2'Min}),
    which follows as a consequence from the previous relations for a vanishing inverse squared
    AdS$_d$-space radius~$r$.

On a programming level, we have solved the third problem of the
paper by means of finding an explicit formalized representation
for the superalgebra $\mathcal{A}^{\prime}(Y(1),AdS_d)$ in terms
of a so-called \textbf{formal  setting of the algorithm}, which
translates the results of the Verma module $V_{\mathcal{A}'}$
realization over a Heisenberg--Weyl superalgebra in a set of
formalized relations (\ref{anticomm})--(\ref{ruleorder}). It is
the relations which, together with the multiplication
table~\ref{table'} and the explicit form of the basis operators of
the superalgebra $\mathcal{A}^{\prime}(Y(1),AdS_d)$
(\ref{t1'+})--(\ref{l2'}), have become the main relations to
realize the programming data model in the language C\# within the
symbolic computation approach.

We have suggested a two-level program model which permits one to
realize, on a programming level, all the properties of an
arbitrary superalgebra of polynomials with an associative
multiplication law as a \emph{basic model of superalgebra}, and
those of proper superalgebra of polynomials from
$\mathcal{A}^{\prime}(Y(1),AdS_d)$ (restricted by the value of
exponent $q$ in $r^q$) as a \emph{polynomial superalgebra model}.
It is shown that in order to describe, in the programming language
C\#, an arbitrary polynomial of finite power in $r$, it is
sufficient to use five basic classes \verb"Expression",
\verb"Coefficient", \verb"Literal", \verb"Product" and \verb"Sum"
from the first level and one class \verb"PhysEnvironment" from the
second level, that is illustrated by Figure~\ref{diarram1}.

We have developed, on a basis of a two-level programming model,
a computer program in C\#, whose main window is shown by
Figure~\ref{figure1}, and which verifies the fact that
the operators of the superalgebra $\mathcal{A}^{\prime}(Y(1),AdS_d)$
satisfy the given algebraic supercommutator relations
by means of a restricted induction principle with
a parameter being the exponent of the inverse squared radius $r$
of the AdS$_d$-space. The validity of the multiplication
table~\ref{table'} is established up to the fourth power in $r$,
which is due to the cyclic character of definitions of the
operators $\mathcal{A}^{\prime}(Y(1),AdS_d)$ in the powers of $r$
practically guarantees the solution of the verification problem
for $q\geq 5$ in $r^q$.

The algorithm, basic data structures, the methods of their
processing and the solution of the formalized problem compose
the basic results of this part of the paper.

Among possible perspectives of research within algebraic and
symbolic computations, we note the problems of constructing
Verma modules and their oscillator realizations for more
involved non-linear algebras and superalgebras corresponding
to higher-spin fields in the AdS$_d$-space subject to a
multi-row Young tableaux, which were discussed in
Ref.~\cite{Resh08122329} for the algebra $\mathcal{A}'(Y(2),AdS_d)$.
This will be by the purpose of a forthcoming work \cite {ReshY2}.
Of course, a detailed verification of the validity of
the corresponding multiplication table of the resulting
expressions for operators of those (super)algebras within
the symbolic computations approach will be a topical problem as well.

As to the development of the program \emph{PhysProject},
one may specify some directions. First of all, it is an
improvement of the visual presentation of data. Second,
the nearest way to enhance the program code of the existing
program model is the swap-out of the second level of data
model and a distribution of the methods to new classes with
respect to those of the first-level model, or an inheritance of
the latter classes and an accumulation of methods.

The general direction of an enhancement of the program consists
in the increasing of its universality in order to adapt the
application of the program to other non-linear algebraic
structures. To these items one may relate a standardization
of the declaration of explicit forms of basis elements such as
$o'_I$, and a definition of multiplication tables, of the rules
for commutation relations. This will permit one to apply the program
to more involved non-linear algebras and superalgebras and
resolve the problem of attaching the program to concrete
superalgebras.

Finally, it is worth noting that our program is assigned to work
with more general objects then $GR$-algebras and corresponding
Gr\"{o}bner bases (see Refs.~\cite{Grobner1,Grobner2} and
references therein)\footnote{Really, the main difference here is
in the fact that the definition of $G$-algebra $\mathcal{A}$ over
field $\mathbb{K}$ \cite{Galgebra}: $A=\mathbb{K}\langle
x_1,...,x_n| \{x_j x_i = c_{ij}\cdot x_ix_j +d_{ij} \}, 1\leq i<j
\leq n\rangle$, $c_{ij}\in \mathbb{K} \setminus\{0\}$ with
$d_{ij}\in \mathbb{K}[x_1,...,x_n]$ of lesser degree than $x_ix_j$
as polynomial, does not provide the realization for
Heisenberg-Weyl superalgebra, i.e. for $d_{ij}=0, c_{ij}=\pm 1$ we
can not realize the relations like: $x_ix_i = 0$ for odd elements
as in (\ref{nilpotf}) due to strict inequality above: $i<j$.}. At
the same time, it is interesting to establish a more detailed
correspondence with these structures and corresponding program
systems for their treatment such as \emph{Plural}, system
\emph{OpenXM} \cite{OpenXM}.

\section*{Acknowledgements}
The authors are grateful to M.S. Plyushchay and T. Tanaka for
pointing out on the application of non-linear algebras and
superalgebras and detailed list of corresponding references.
A.A.R. thanks Yu.Zinoviev for discussion of the peculiarities of
the program work, R.R. Metsaev for his advice to find or develop
programming tools for a verification of the validity of commutator
multiplication table for operators given as a formal power series
in the powers of oscillator variables. The work of the program was
demonstrated on the 4th Sakharov's International Conference in
Moscow, May, 18-23, 2009 (see \cite{Sakharov}) and on the
International Workshop "Supersymmetry and Quantum Symmetries"
(SQS'09) in Dubna, July 29 -- August 3, 2009.


\end{document}